\documentclass{elsarticle}
\usepackage{cleveref}

\journal{Nuclear Physics B}
\usepackage{graphicx}
\usepackage{amsmath}
\usepackage{amssymb}
\usepackage{dsfont}
\usepackage{wasysym}
\usepackage{color}
\usepackage{caption}
\usepackage{subfig}
\allowdisplaybreaks[4]

\newcommand{\href}[1]{}
\newcommand{\path}[1]{}

\usepackage{ntimes}
\definecolor{gray}{rgb}{0.5,0.5,0.5}

\renewcommand{\paragraph}[1]{ \par \vspace{0.25cm} {\bf #1}.}

\begin{document}
\begin{frontmatter}
\title{Asymptotic correlation functions and FFLO signature for the one-dimensional attractive Hubbard model}
\author[WIPM,ANU,UCAS]{Song Cheng}
\author[WIPM,CCAP]{Yuzhu Jiang}
\author[WIPM,CCAP]{Yi-Cong Yu}
\author[CU,ANU,ANU-math]{Murray T. Batchelor}
\author[WIPM,CCAP,ANU]{Xi-Wen Guan{\footnote{xiwen.guan@anu.edu.au}} }
\address[WIPM]{State Key Laboratory of Magnetic Resonance and Atomic and Molecular Physics,\\
Wuhan Institute of Physics and Mathematics, Chinese Academy of Sciences, Wuhan 430071, China}
\address[CCAP]{Center for Cold Atom Physics, Chinese Academy of Sciences, Wuhan 430071, China}
\address[CU]{Centre for Modern Physics, Chongqing University, Chongqing 400044, China}
\address[ANU]{Department of Theoretical Physics, Research School of Physics and Engineering,\\
Australian National University, Canberra ACT 0200, Australia}
\address[ANU-math]{Mathematical Sciences Institute,
Australian National University, Canberra ACT 0200, Australia}
\address[UCAS]{University of Chinese Academy of Sciences, Beijing 100049, China}
\begin{abstract}
We study the long-distance asymptotic behavior of various correlation functions for the one-dimensional (1D) attractive Hubbard model
in a partially polarized phase through the Bethe ansatz and conformal field theory approaches.
We particularly  find the oscillating behavior of these correlation functions with spatial power-law decay, of
which the pair (spin) correlation function oscillates with a frequency $\Delta k_F$ ($2\Delta k_F$). Here $\Delta k_F=\pi(n_\uparrow-n_\downarrow)$
is the mismatch in the Fermi surfaces of spin-up and spin-down particles.
Consequently, the pair correlation function in momentum space has peaks at the mismatch $k=\Delta k_F$, which has been observed in recent numerical work on this model.
These singular peaks in momentum space together with the spatial oscillation suggest an analog of the Fulde-Ferrell-Larkin-Ovchinnikov (FFLO) state in  the 1D Hubbard model.
The  parameter $\beta$ representing the lattice effect becomes  prominent  in  critical exponents  which determine the power-law decay of all correlation functions.
We point out that the backscattering of unpaired fermions and bound pairs within their own Fermi points gives  a microscopic origin of  the FFLO  pairing in 1D.

\end{abstract}


\end{frontmatter}
\section{Introduction}

With addition of an on-site interaction term to the tight binding hamiltonian,
the Hubbard model successfully provides a paradigm for condensed matter physics \cite{Rase,Monto,Baer,Ess}.
In contrast to its simple form, this model exhibits diverse features of many-body systems,
such as a Mott phase, high $T_c$ superconductivity, quantum phase transition, Fulde-Ferrell-Larkin-Ovchinnikov (FFLO) phase, spin-charge separation etc.
Recently this model has received further attention in the experimental developments of trapping ultracold atoms on optical lattices,
which provides a prominent opportunity to verify theoretical predictions of this kind \cite{Singha2011,Hart2015,Greif2016,Parsons2016S,Cheuk2016S,Boll2016S}.
The one-dimensional (1D) Hubbard model can be exactly solved through the application of the Bethe ansatz \cite{Lieb1968PRL,Lieb2003PA},
a thorough study of which may also help to understand key aspects of many-body physics in higher dimensions.

The Hubbard model with attractive interaction is considered as a promising candidate to explain high $T_c$ superconductivity.
To this end, understanding the pairing mechanism in 1D is of significant importance.
According to the Bardeen-Cooper-Schrieffer (BCS) theory, a Cooper pair is formed by electrons with opposite spins and momenta and total momentum zero.
This balance between Fermi energies breaks down in the presence of strong magnetic field, so that a novel superconductive state -- the FFLO state -- appears \cite{FF1964,LO1965}.
The FFLO pair carries non-zero centre-of-mass momentum, originating from the strong magnetic field.
The superconducting order parameter and density of spins in the FFLO state exhibit a periodic oscillation in the spatial coordinate.
The experimental observation of the FFLO state in various materials has been sought for decades.
Within this scenario, more evidence has been found in heavy-fermion systems \cite{Matsuda2007,Ptok2017}.

The pair mechanism of the 1D attractive Hubbard model has been investigated \cite{Bogoliubov1988,Bogoliubov1989,Bogoliubov1990}.
Cooper pairs in 1D exist in the region $B<B_{c1}$ below the critical magnetic field $B_{c1}$, where the average distance between pairs is much larger than the average pair size.
This means that the single-particle Green's function decays exponentially and the single-pair correlation function decays as a power of distance.
Once exceeding the critical magnetic field ($B>B_{c1}$), the field begins to break up the pairs, and both the above correlation functions decay as a power of distance.

The FFLO state in the 1D attractive Hubbard model has been extensively  studied by various methods, such as the numerical approaches of the density matrix renormalization group (DMRG) \cite{Feiguin2007,Luscher2008,Rizz2008,Tezuka2008,Tezuka2010} and quantum Monte Carlo (QMC) \cite{Batrouni2008,Wolak2010}.
It has been found that the pair correlation complies with a power-law decay, i.e., $n_{pair} \varpropto \cos \left( k_{FFLO} |x| \right)/|x|^\alpha$,
and its corresponding momentum distribution has peaks in the position of $k_{FFLO}=\pi\left( n_\uparrow - n_\downarrow \right)$ \cite{Feiguin2007}.
As far as we know,  the confirmation of the FFLO state in 1D mainly relies on numerics, although the application of conformal field theory (CFT)
has provided a definite answer for the FFLO signature of the 1D attractive Fermi gas \cite{Lee2011}.
In this paper we focus on asymptotic correlation functions  of the 1D attractive Hubbard model in the partially polarized phase.
In particular, we investigate the long-distance asymptotics of the single-particle Green's function, the charge density correlation function, the spin correlation function and the pair correlation function.
We present  a theoretical confirmation of the existence of the FFLO state in the 1D attractive Hubbard model.
In contrast to  the continuum case of the 1D attractive SU(2) Fermi gas, here the lattice effects characterized by a parameter $\beta$ become prominent
in the critical exponents of all correlation functions.

The paper is organized as follows. In Sec. \ref{sec-model}, we make a brief introduction to the exact solution of the 1D attractive Hubbard model and
derive the finite-size correction to the ground state. In Sec. \ref{LLE}, we discuss three types of elementary excitations and the dressed charge integral equations.
The dressed charge matrix is derived in the low density limit, which affords access to the conformal dimensions.
The long-distance asymptotics of the various correlation functions are studied in Sec. \ref{sec-CR}, along with the discussion of the FFLO signature in the partially polarized phase.
Sec. \ref{sec-c} is reserved for the conclusion.

\section{Exact solution of the Hubbard model and ground state}
\label{sec-model}
The 1D fermionic Hubbard model is described by the hamiltonian
\begin{align}
 \label{hamiltonian}
 H=&-\sum_{j=1}^L \sum_{a=\uparrow,\downarrow} \left( c_{j,a}^\dagger c_{j+1,a} + c_{j+1,a}^\dagger c_{j,a} \right) + u \sum_{j=1}^L \left( 1-2n_{j,\uparrow} \right) \left( 1-2n_{j,\downarrow} \right) \nonumber\\
 &-B \sum_{j=1}^L (n_{\uparrow,j}-n_{\downarrow,j})-\mu\sum_{j=1}^L (n_{\uparrow,j}+n_{\downarrow,j}),
\end{align}
where $\hat c_{j,a}^\dagger$ and $\hat c_{j,a}$, with $\hat n_{j,a}=\hat c_{j,a}^\dagger \hat c_{j,a}$,
are the creation and annihilation operators of fermions  with spin $a$ ($a= \, \uparrow$ or $a=\, \downarrow$) at site $j$ on a 1D lattice of length $L$.
 The chemical potential and magnetic field are denoted by $\mu$ and $B$, respectively.
 Meanwhile $u$ represents the on-site  interaction between particles ($u>0$ for repulsion and $u<0$ for attraction).
The hamiltonian (\ref{hamiltonian}) is exactly solvable with periodic or open boundary
conditions \cite{Ess,Lieb1968PRL,Lieb2003PA,Guan:2000,Zhou:1996,Guan:1997,Shiroishi:1997,Li:2014}.

By means of the (nested) Bethe ansatz, the diagonalization of hamiltonian (\ref{hamiltonian}) leads to a set of nonlinear algebraic equations,
known as the Lieb-Wu equations and written as \cite{Lieb1968PRL,Lieb2003PA}
\begin{align}
 & \exp({\rm i}  k_j L)=\prod_{\alpha=1}^M \frac{\sin k_j-\Lambda_\alpha+{\rm i}  u}{\sin k_j-\Lambda_\alpha-{\rm i}  u},\hspace{58pt} j=1, 2, \ldots,N, \label{lw1}\\
 &
 \prod_{j=1}^N \frac{\sin k_j-\Lambda_\beta+{\rm i}  u}{\sin k_j-\Lambda_\beta-{\rm i}  u}
 =-\prod_{\alpha=1}^M \frac{\Lambda_\alpha-\Lambda_\beta+2 {\rm i}  u}{\Lambda_\alpha-\Lambda_\beta-2 {\rm i}  u},\quad \beta=1, 2, \ldots,M. \label{lw2}
\end{align}
Here $\Lambda$ is the spin rapidity and $k$ is the quasimomenta of fermions.
$N$ and $M$ are the total number of fermions  and the number of fermions  with down spins.
The energy $E$ and momentum $P$ of this model are
\begin{align}
E=&-2\sum_{j=1}^N \cos k_j+u(L-2N)-2Bm-\mu N,\label{energy}\\
P=& \sum_{j=1}^N k_j \mod 2\pi.  \label{momentum}
\end{align}
The magnetization per site $m=\frac{N-2M}{2L}$.

The distributions of the quasimomenta $\left\{ k_i \right\}$ with $i=1,2, \ldots, N$
and spin rapidities $\left\{ \Lambda_\beta \right\}$ with $\beta =1,2, \ldots, M$ in the thermodynamic limit comply with the string hypothesis.
Accordingly, all roots of the Lieb-Wu equations are divided into three categories.
These are single real $k$, $k$-$\Lambda$ strings and $\Lambda$-$\Lambda$ strings \cite{Lee1988,Ess1994b,SC}.
Namely single real $k$'s,  the $\alpha$-th $k$-$\Lambda$ string of length $\boldsymbol{m}$, with
\begin{align}\label{string-hyp1}
k_\alpha^1&=\arcsin({\Lambda_\alpha^\prime}^m+ \mathrm{i} \, m \, |u|),\nonumber \\
k_\alpha^2&=\arcsin({\Lambda_\alpha^\prime}^m+ \mathrm{i} \,  (m-2) \,  |u|),\nonumber \\
k_\alpha^3&=\pi-k_\alpha^2,\nonumber \\
\vdots \nonumber\\
k_\alpha^{2m-2}&=\arcsin({\Lambda_\alpha^\prime}^m- \mathrm{i} \,(m-2)\, |u|),\nonumber \\
k_\alpha^{2m}&=\arcsin({\Lambda_\alpha^\prime}^m-\mathrm{i} \, m \, |u|),
\end{align}
for which there are  $\boldsymbol{2m}$ $k$'s accompanied by $\boldsymbol{m}$ spin-rapidities  in  $\Lambda$ space, with
\begin{eqnarray}
{\Lambda_\alpha^\prime}^{m,j}&=&{\Lambda_\alpha^\prime}^m+ \mathrm{i} \, (m+1-2j) \,  |u|,\label{string-hyp2}
\end{eqnarray}
 where  $j=1,2,\ldots, m$ and ${\Lambda_\alpha^\prime}^m$ is the real center of the $k-\Lambda$ string.
There is also the $\beta$-th $\Lambda$-$\Lambda$ string of length $\boldsymbol{m}$
  \begin{equation}\label{string-hyp3}
  \Lambda_\beta^{m,j}=\Lambda_\beta^m+ \mathrm{i} \, (m+1-2j)\, |u|,
  \end{equation}
where $ j=1,2, \ldots, m,$ and $\Lambda_\beta^m$ is the real center of the $\Lambda$ string.
The $\Lambda$ strings represent the spin wave bound states in the spin sector.

The situation for the ground state is much simplified, where only single real $k$'s and $k$-$\Lambda$ strings of length one ($m=1$) are permitted.
The $\Lambda$-$\Lambda$ strings are suppressed due to the ferromagnetic ordering,  see analysis in detail in \cite{SC}.
This means that the quasimomenta of fermions in bound pairs and the excess unpaired fermions can be respectively written
as $k^b_{\eta,\pm}= \arcsin(k^{b}_\gamma \pm{\rm i} |u|)$ and $k^u_\beta$, with $ \eta=1,2, \ldots, M$.
For simplicity,  we have denoted $k$ as $k^u_\beta$ and $\Lambda$ as $k^b_\gamma$.
Substituting the above simplified string hypothesis into Eqs. (\ref{lw1}) and (\ref{lw2}) and then taking logarithms yields the discrete Bethe ansatz equations in the form
\begin{align}
 & \label{qn-u}
 \frac{2\pi}L I_j^{u} = k_j^{u} -\frac{1}{L} \sum_{\alpha=1}^{N^{b}} \theta\left( \frac{\sin k^{u}_j-k_\alpha^{b}}{|u|} \right),
 \\
 &\label{qn-b}
 \frac{2\pi}L  I^{b}_\alpha =
 2{\,\rm Re} \left[ \arcsin(k^{b}_\alpha+{\rm i} |u|) \right]
 -\frac{1}{L} \sum_{j=1}^{N^{u}} \theta \left( \frac{k^{b}_\alpha - \sin k_j^{u}}{|u|} \right) - \frac{1}{L} \sum_{\beta=1}^{N^{b}} \theta \left( \frac{ k^{b}_\alpha-k^{b}_\beta}{2|u|} \right),
\end{align}
where $N^u$ ($N^b$) is the number of unpaired fermions (bound pairs) satisfying $N^u+2N^b=N$.  The quantum numbers $I_j^{u}$ and $I_\alpha^{b}$ take the values
\begin{align}
 \label{qn-g}
 I_j^{u} \in \mathbb{Z}+\frac{N^{b}}{2}, \quad I_\alpha^{b} \in \mathbb{Z} + \frac{N^{u} + N^{b}+1}{2}.
\end{align}
The ground state energy and momentum are explicitly expressed as
\begin{align}
E=&\sum_{j=1}^{N^u} \left( -2\cos k_j^u - \mu-2u-B \right) + \sum_{\gamma=1}^{N^b} \left[ -2\mu-4u- 4 \textmd{Re}\sqrt{1-\left( k_\gamma^b+\mathrm{i}|u| \right)^2} \right],\\
P=&\sum_{j=1}^{N^u} k_j^u + \sum_{\eta}^{N^b} \left(k_{\eta,+}^b + k_{\eta,-}^b  \right).
\end{align}

We define counting functions, $y_L^u(k_j^u)={2\pi I_j^u}/{L}$ and $y_L^b(k_\beta^b)={2\pi I_\beta^b}/{L}$, which are monotonic increasing functions and satisfy the equations
\begin{align}
\rho_L^u(k^u)=&\frac{1}{2\pi}\frac{\textmd{d}y_L^u(k^u)}{\textmd{d}k^u}\notag \\
=&\frac{1}{2\pi}-\frac{1}{L} \sum_{j=1}^{N^b} a_1\left( \sin k^u - k_j^b \right)  \cos k^u, \\
\rho_L^b(k^b)=&\frac{1}{2\pi} \frac{\textmd{d}y_L^b(k^b)}{\textmd{d}k^b}\notag \\
=&\frac{1}{2\pi} \int_{-\pi}^\pi \textmd{d}k \, a_1(k^b-\sin k) - \frac{1}{L} \sum_{j=1}^{N^u} a_1(k^b-\sin k_j^u) - \frac{1}{L}\sum_{j=1}^{N^b} a_2(k^b-k_j^b).
\end{align}
Here $\rho_L^\gamma(k^\gamma)$ ($\gamma=u,b$) is the root density of the corresponding quasimomentum and $a_n(x)=\frac{1}{\pi} \frac{n|u|}{(nu)^2+x^2}$.

To obtain finite-size corrections in terms of the  above root densities when $L \gg 1$, we utilize the Euler-MacLaurin formula to obtain the integral equations (up to high orders)
\begin{eqnarray}
\rho_L^u(k^u)
&=&\frac{1}{2\pi}-\cos k^u \int_{Q_-^b}^{Q_+^b} \textmd{d}k^b \, a_1(\sin k^u-k^b) \rho_L^b(k^b) \nonumber\\
&&-\frac{1}{24 L^2} \cos k^u \left. \frac{a_1^\prime(\sin k^u-k^b)}{\rho_L^b(k^b)} \right|_{k^b=Q_-^b}^{k_b=Q_+^b} \label{rho-L-u}\\
\rho_L^b(k^b)
&=&\frac{1}{2\pi} \int_{-\pi}^\pi \textmd{d}k \, a_1(k^b-\sin k) - \int_{Q_-^u}^{Q_+^u} \textmd{d}k^u \, a_1(k^b-\sin k^u) \rho_L^u(k^u) \nonumber \\
&& - \int_{Q_-^b}^{Q_+^b} \textmd{d}k \, a_2(k^b-k) \rho_L^b(k)   \label{rho-L-b}\\
&& -\frac{1}{24 L^2} \left[ \left. \frac{\cos k^u \, a_1^\prime(k^b-\sin(k^u))}{\rho^u_L(k^u)}\right|_{k^u=Q_-^u}^{k^u=Q_+^u}
+ \left. \frac{a_2^\prime(k^b-k)}{\rho^b_L(k)} \right|_{k=Q_-^b}^{k=Q_+^b} \right],\nonumber
\end{eqnarray}
where $a_n^\prime(x)$ is the derivative of $a_n(x)$ and $Q^\gamma_\pm$ ($\gamma=u,b$) denote the Fermi points.

\begin{figure}[ht]
 \begin{center}
 \includegraphics[width=1.0\linewidth]{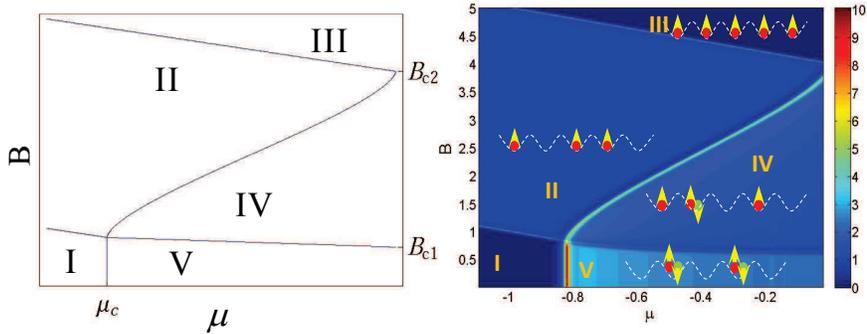}
 \end{center}
 \caption{The phase diagram of the  1D attractive Hubbard model. The left panel  is for $T=0$,
 where $\mu_c=2|u|-\sqrt{u^2+1}$, $B_{c1}=2|u|-2+2\int_{-\infty}^\infty \textmd{d}w \frac{J_1(w)\exp(-|u|w)}{w\cosh(uw)}$, and $B_{c2}=2+2|u|$, see analytical phase boundaries in \cite{SC}.
 The right panel is a contour  plot of the compressibility Wilson ratio $R_W^\kappa$ with interaction $u=-1$ and temperature $T=10^{-3}$.
 The values of the Wilson ratio have a sudden enhancements in the vicinities of the phase boundaries at low  temperatures.
 Both diagrams coincide and  present  five quantum phases. These are I: vacuum; II: fully polarized phase, less than half-filling;
 III:   half-filling phase and  fully polarized; IV: partially polarized phase, less than half-filling; V: fully paired phase and less than half-filling.}
 \label{f-pd}
\end{figure}

It is necessary to introduce the thermodynamic Bethe ansatz (TBA) equations for the derivation of finite-size corrections to the ground state and low-lying excitations.
The TBA equations  describe the full  thermodynamics of the model in the whole  temperature regime \cite{Lee1988,Ess1994b,SC}.
Building on these equations,  the equation of states can be  derived in terms of densities of single fermions and bound states \cite{SC}.
In the ground state, the phase diagram consists of five states of  the bound pairs  and excess  fermions in the $B$-$\mu$ plane,  see Fig.~\ref{f-pd}.
The $T=0$ phase diagram is presented in the left panel of Fig.~\ref{f-pd}, whereas the right panel presents the low temperature phase diagram obtained via
 the compressibility Wilson ratio, which is defined by $R_W^\kappa=\frac{\pi^2 k_B^2}{3} \frac{\kappa}{C_v/T}$ \cite{SC} in terms of the
 Boltzmann constant $k_B$, the compressibility $\kappa$ and the specific heat $C_v$.
 This dimensionless ratio measures the competition between quantum and thermal fluctuations,
 which exhibits an enhancement near the quantum critical points and thus serves as a powerful tool in determining the low temperature phase diagram.
 Hereafter we only concentrate on the partially polarized  phase IV, which, at zero temperature, presents a novel FFLO-like state on a 1D lattice.
 At low temperature,  the low energy physics of this phase reveals a subtle  two-component Tomonaga-Luttinger liquid (TLL),  composed of bound pairs and of unpaired fermions.

The TBA equations for the ground state of the 1D attractive Hubbard model are written as
\begin{align}
\varepsilon^u \left(k^u\right)=&-2\cos k^u-\mu-2u-B-\int_{Q^b_-}^{Q^b_+} \textmd{d}k^b \, a_1(\sin k^u-k^b) \varepsilon^b (k^b)\label{new0tba1}\\
\varepsilon^b \left(k^b\right)=&-2\mu-2\int_{-\pi}^\pi \textmd{d}k^u \, \cos^2 k^u\, a_1(\sin k^u-k^b)
- \int_{Q^b_-}^{Q^b_+} \textmd{d}\Lambda \, a_2(k^b-\Lambda)\varepsilon^b(\Lambda)\notag \\
&-\int_{Q^u_-}^{Q^u_+} \textmd{d}k^u \, \cos k^u \, a_1(\sin k^u-k^b) \varepsilon^u (k^u).
\label{new0tba2}
\end{align}
These are also called the dressed energy equations  $\varepsilon^{b,u}(k^{b,u})$ describing the excitations of bound pairs and unpaired fermions for pure charge degrees of freedom.
In contrast to the low-lying excitations in the repulsive Hubbard model \cite{Ess}, the excess fermions cause a ferromagnetic fluctuation which was  fully  suppressed  in zero  temperature limit, see analysis in \cite{SC}.
Therefore the only  existence of  the Fermi points for the bound pairs and unpaired fermions results in  two branches of gapless excitations in low energy sector.
Thus the calculation of finite-size correction  is  naturally accessible by virtue of root density formalism.
For the pure paired phase of this model, a similar study was carried out in \cite{Bogoliubov1989}.

The conformal invariance of a 1D many-body system at $T=0$ provides a universality class of criticality in terms of the central charge  $C$ of the underlying Virasoro algebra.
Indeed, the dimensionless central charge classifies the finite-size scaling form of energies in low-lying excitations.
In particular, the $C=1$ universality class gives rise to a systematic calculation of the critical exponents which govern the power-law decay of correlation functions in long-distance \cite{Belavin1984,Blote1986,Affleck1986}.
With the help of the root densities presented in Eqs. (\ref{rho-L-u}) and (\ref{rho-L-b}) and the TBA equations (\ref{new0tba1}) and (\ref{new0tba2}),
we derive the finite-size correction to the ground state energy in the form
\begin{align}
 \frac{\Delta E}{L} = -\sum_{\gamma={u,b}} \frac{C \pi}{6 L^2} v^\gamma,\label{finite-s-c-energy-gr}
\end{align}
where $C=1$ is the central charge for  both branches of excitations.
In the above equations $v^{u,b}$ are the sound velocities of the unpaired fermions and bound pairs.
The sound velocities  are  defined by $v^\gamma=\left. \pm \frac{\textmd{d} \varepsilon^\gamma(k^\gamma)}{\textmd{d}p^\gamma(k^\gamma)}\right|_{k^\gamma= \pm Q^\gamma}=\left.\pm \frac{\textmd{d}\varepsilon^\gamma/\textmd{d}k^\gamma}{2\pi \rho^\gamma(k^\gamma)}\right|_{k^\gamma= \pm Q^\gamma}$ for $\gamma=u,b$.
Here we have denoted  $\pm Q^\gamma$ as  the Fermi points of corresponding single fermions and bound pairs and the momenta  $p^\gamma(k^\gamma)=\lim_{L\rightarrow \infty} y^\gamma_L(k^\gamma)$.

\section{Low-lying excitations and dressed charge matrix}
\label{LLE}

In order to obtain the conformal dimensions and the critical exponents, the finite-size corrections to the low-lying excitations need to be calculated.
The 1D attractive Hubbard model has two branches of gapless excitations in terms of the unpaired fermions and bound pairs at zero temperature.
The method used in this work follows the scheme established in Refs. \cite{Ess,Woyna1989,Frahm1990PRB,Frahm1991PRB}.
In general, the low-lying excitations can be realized by combination of three types of elementary excitations,
all of which involve the distortion of the Fermi points due to  the transformation or variation of the Fermi points.
Such changes  can be characterized  by the changes in  the quantum numbers given in Eq. (\ref{qn-g}).

\begin{figure}[ht]
\begin{center}
\includegraphics[width=1.0\linewidth]{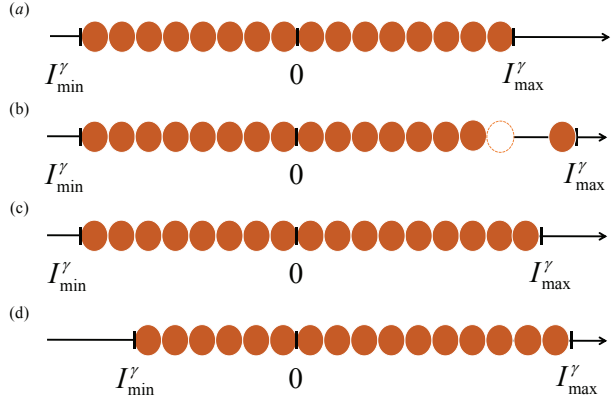}
\end{center}
\caption{A sketch of distributions for the quantum number $I_j^\gamma$ ($\gamma=u,b$) in the ground state and excited states.
There are three types of elementary excitations:  (a)  the symmetric distribution for the  ground state;  (b) a particle-hole excitation near the right Fermi point, the  Type I excitation;
(c) one more particle is added  to the right Fermi point, the Type II excitation;  (d) two particles near the left Fermi point are moved to the right Fermi point, the Type III excitation.
One may find that all of the possible vacancies  for quantum numbers in Type II and Type III  excitations are occupied by particles, which  are  different from the Type I  excitation.}
\label{f-et}
\end{figure}

Excitations of Type I move particles inside the Fermi sea to location $j$  outside the Fermi sea,  known as a  particle-hole excitation, see Fig.~\ref{f-et}~(b).
The lowest particle-hole excitation is  described by the change of quantum numbers
$I^\gamma_{j}$ close to $I^\gamma_\pm$ ($\gamma=u,b$), where $I^\gamma_+=I^\gamma_{\mathrm{max}}+\frac{1}{2}$ and $I^\gamma_-=I^\gamma_{\mathrm{min}}-\frac{1}{2}$.
Here  the superscript $u$ stands for  the unpaired fermions and $b$ for the bound pairs, respectively.
The changes of the total momentum and energy  are given by
\begin{align}
 \Delta P=&\frac{2\pi}{L}\sum_{\gamma={u,b}} \big(N^{\gamma}_+- N^{\gamma}_-\big)  \mod 2\pi,  \label{elem-exci-1-momentum}\\
 \Delta E=&\sum_{\gamma={u,b}} \frac{2\pi}{L} v^\gamma \big( N_+^\gamma+N_-^\gamma \big), \label{elem-exci-1-energy}
\end{align}
where $N^\gamma_+>0$ ($N^\gamma_->0$) stands for the change of the quantum number for corresponding particle near the right (left) Fermi point.

Type II excitations originate from the change of particle numbers of   unpaired fermions and bound pairs in  the Fermi seas, see Fig.~\ref{f-et}~(c).
It is easy to see that the particle number is given by $N^\gamma=I^\gamma_+-I^\gamma_-$,
and a Type II excitation is characterized by quantum number $\Delta N^\gamma=N^\gamma_e-N^\gamma_g$ where subscripts $e$ and $g$ respectively represent the excited state and ground state.

Type III  excitations are caused by the backscattering process, where  particles from one Fermi point move  to the other one, see Fig.~\ref{f-et}~(d).
They are characterized by quantum number $\Delta D^\gamma=\frac{I^\gamma_+ + I^\gamma_-}{2}$, ($\gamma=u,b$).
In light of the `continuous' Fermi sea in the Type II and Type III  excitations, quantum number $\Delta N^\alpha$ and $\Delta D^\alpha$
characterize the changes of  the sizes of Fermi seas and the displacements of the center of Fermi seas,  respectively.

The calculation of finite-size corrections for Type I excitations is straightforward.
Calculations for the Type II and Type III excitations can also be carried out in a systematic way.
These calculations are given in the Appendix.
Here we summarize the results for the three types of elementary excitations.
For convenience in the following calculation of the conformal dimensions,
the excitations  can be cast   into  unified finite-size scaling forms of the energy and total momentum, which read
\begin{align}
\Delta E
=&\frac{2\pi}{L}\bigg[ \frac{1}{4} \left( \Delta\vec{N} \right)^t \cdot \left( \hat{Z}^{-1} \right)^t \cdot \hat{S}_v \cdot \hat{Z}^{-1} \cdot \Delta\vec{N} + \left( \Delta\vec{D} \right)^t \cdot \hat{Z} \cdot \hat{S}_v \cdot \hat{Z}^{t} \cdot \Delta\vec{D} \notag\\
& + v^\alpha \left(N_+^\alpha+N_-^\alpha \right) \bigg], \label{finite-s-c-energy-ex} \\
\Delta P=& \frac{2\pi}{L} \sum_{\alpha=u,b} \Delta D^\alpha \cdot \left[ N_+^\alpha - N_-^\alpha + \Delta D^\alpha \left( N^\alpha+\Delta N^\alpha \right) \right].\label{finite-s-c-momentum-ex}
\end{align}
For both results there are additional higher order corrections.
In these equations we have introduced the notation
\begin{align}
\Delta \vec{N}=\left[
                 \begin{array}{c}
                   \Delta N^u \\
                   \Delta N^b \\
                 \end{array}
               \right],
\quad
\Delta \vec{D}=\left[
                 \begin{array}{c}
                   \Delta D^u \\
                   \Delta D^b \\
                 \end{array}
               \right],
\end{align}
\begin{align}
\hat{S}_v=\left[
            \begin{array}{cc}
              v^u & 0 \\
              0 & v^b \\
            \end{array}
          \right],
\quad
\hat{Z}=&\left.\left[
              \begin{array}{cc}
                Z_{uu}(k^u=Q_+^u) & Z_{ub}(k^b=Q_+^b) \\
                Z_{bu}(k^u=Q_+^u) & Z_{bb}(k^b=Q_+^b) \\
              \end{array}
            \right]\right|_G.
\end{align}
The  superscript $t$ in the above equations stands for matrix transpose, and the quantum numbers $\Delta D^\beta$ ($\beta=u,b$) obey the relations
\begin{align}
\Delta D^u \in \frac{\Delta N^u + \Delta N^b}{2} + \mathds{Z}, \quad \Delta D^b \in \frac{\Delta N^u}{2} +\mathds{Z}, \label{choosing-relation}
\end{align}
which stems from the Type III excitations with  the quantum numbers $I^\beta_\pm$ ($\beta=u,b$)  through the relation  Eq. (\ref{qn-g}).
Although the parities of particle numbers $N^u$ and $N^b$ do not affect  the physical properties of the model  in the thermodynamic limit,
  they may  affect  the degeneracy of the ground state and also affect the parities of $\Delta D^{u,b}$.
For example, when both $N_g^u$ and $N_g^b$ are even numbers, according to the first equation of (\ref{qn-g}), the quantum numbers $I^u$ of the ground state have two possible choices, $\{-N_g^u/2,-N_g^u/2+1,\cdots, N_g^u/2-1\}$ and $\{-N_g^u/2+1,-N_g^u/2+2,\cdots, N_g^u/2\}$.
In this sense, the ground state is  two-folder degeneracy.
On the other hand, when $N^b_g$ is odd and $N^u_g$ is even, the ground state is nondegenerate.
In the following discussion on  the correlation functions, without loss generality, we will choose $N^b_g$ to be odd and  $N^u_g$ to be even.

To obtain  the relation (\ref{choosing-relation}), one may use the definitions of $\Delta D^\beta$ and $I^\beta_\pm$.  It is easily seen that $\Delta D^u=\frac{I_+^u-I_-^u}{2}+I_-^u$, where $I_\pm^u$ are the quantum numbers of excited states and $I_-^u=I_1^u-\frac{1}{2}$ is the left most  quantum number, its parity  depends on $N^b$ through the first equation of (\ref{qn-g}). Based on the supposed parities of $N^b_g$  and $N^u_g$, thus we have
\begin{align}
\Delta D^u &\in \frac{\Delta N^u + N^u_g}{2}  + \frac{\Delta N^b+N^b_g-1}{2} +\mathds{Z}\notag\\
&\in \frac{\Delta N^u+\Delta N^b}{2} +\mathds{Z}. \notag
\end{align}
Similarly,  one can derive the result for $\Delta D^b$.

The dressed charges at the Fermi points $Q_+^{\gamma}$ ($\gamma=u,b$) are obtained from the elements of the dressed charge matrix, which satisfy the  integral equations
\begin{eqnarray}
Z_{uu}(k^u)&=&\,1-\int_{Q^b_-}^{Q^b_+} \textmd{d}k^b \, a_1(k^b-\sin k^u) \, Z_{ub}(k^b), \label{dress-charge-equation-1} \\
Z_{ub}(k^b)&=&- \int_{Q^u_-}^{Q^u_+} \textmd{d}k^u \, \cos k^u \, a_1(\sin k^u-k^b) \, Z_{uu}(k^u) \nonumber\\
&&- \int_{Q^b_-}^{Q^b_+} \textmd{d}\tilde{k}^b \, a_2(k^b-\tilde{k}^b) \, Z_{ub}(\tilde{k}^b), \label{dress-charge-equation-2} \\
Z_{bu}(k^u)&=&-\int_{Q^b_-}^{Q^b_+} \textmd{d}k^b \, a_1(k^b-\sin k^u)\, Z_{bb}(k^b), \label{dress-charge-equation-3} \\
Z_{bb}(k^b)&=&\,1-\int_{Q^u_-}^{Q^u_+} \textmd{d}k^u \, \cos k^u \, a_1(\sin k^u-k^b)\, Z_{bu}(k^u) \nonumber\\
&&- \int_{Q^b_-}^{Q^b_+} \textmd{d}\tilde{k}^b \, a_2(k^b-\tilde{k}^b) \, Z_{bb}(\tilde{k}^b),\label{dress-charge-equation-4}
\end{eqnarray}
which were also given  from different aspect \cite{Zvyagin:1993}.
We note that the form of the dressed charges is quite different from those for the 1D repulsive Hubbard model \cite{Ess}.

In the ground state, phase V is gapped in the spin sector due to the existence of the  bound pairs.
However, the system  becomes gapless if the magnetic field is greater than the lower critical field,  at which  bound pairs break.
Consequently, phase IV consists of both bound pairs and  unpaired fermions.
The  conformal invariant symmetry enables one to obtain the finite-size scaling forms Eqs. (\ref{finite-s-c-energy-gr}) and (\ref{finite-s-c-energy-ex}).
In what follows we will calculate the conformal dimensions which determine the critical exponents of two-point correlation functions between primary fields $\langle \hat{O}^\dag(x,t) \hat{O}(x',t') \rangle$.

We focus on the asymptotics of correlation functions in phase V.
For this phase,  one expects a  power-law decay  of the correlation function at  $T=0$.
Meanwhile, at   $T>0$,  the correlation functions should decay  exponentially.
The conformal scaling dimensions can be read off as
\begin{align}
2\Delta_\pm^u=&\left( \hat{Z}_{uu} \cdot \Delta D^u + \hat{Z}_{bu} \cdot \Delta D^b \pm \frac{\hat{Z}_{bb} \cdot \Delta N^u - \hat{Z}_{ub} \cdot \Delta N^b}{2\det\{\hat{Z}\}} \right)^2+2N_\pm^u, \label{cfd-1}\\
2\Delta_\pm^b=&\left( \hat{Z}_{ub} \cdot \Delta D^u + \hat{Z}_{bb} \cdot \Delta D^b \pm \frac{\hat{Z}_{uu} \cdot \Delta N^b - \hat{Z}_{bu} \cdot \Delta N^u}{2\det\{\hat{Z}\}} \right)^2+2N_\pm^b,\label{cfd-2}
\end{align}
where $N^\alpha_\pm$ ($\alpha=u,b$) characterizes the descendent field from the primary field.
It follows that the  long-distance asymptotics of the two point correlation functions  are given by
\begin{align}
\langle \hat{O}(x,t) \,\hat{O}(0,0) \rangle=\frac{\exp \left[ -\mathrm{i}  \frac{2\pi}{L} \left( N^u\cdot \Delta D^u + N^b \cdot \Delta D^b \right)x \right]}{(x-\mathrm{i} v^u  t)^{2\Delta_+^u} \, (x+\mathrm{i}  v^u t)^{2\Delta_-^u} \, (x-\mathrm{i}  v^b  t)^{2\Delta_+^b} \, (x+\mathrm{i}  v^b t)^{2\Delta_-^b} }.\label{typical-correlation}
\end{align}

The dressed charge equations can be  simplified in the low density regime, i.e., with small integration boundaries $Q^\gamma \ll 1$ ($\gamma=u,b$).
Here we replace $Q^\gamma_\pm$ by $\pm Q^\gamma$ in the dressed charge matrix, whose elements are calculated for the  ground state.
Obviously, the dressed charge equations can be separated into two sets of coupled  integral equations,
composed of Eqs. (\ref{dress-charge-equation-1}) and (\ref{dress-charge-equation-2}), and of Eqs. (\ref{dress-charge-equation-3}) and (\ref{dress-charge-equation-4}), respectively.
By analysing the order of $Q^\gamma$ in the dressed equations Eqs. (\ref{dress-charge-equation-1}) - (\ref{dress-charge-equation-4})
we can further obtain asymptotic forms of these equations.

To begin,  we substitute Eq. (\ref{dress-charge-equation-2}) into Eq. (\ref{dress-charge-equation-1}) to give
\begin{equation}
Z_{uu}(k^u)\approx  1.
\end{equation}
We further substitute this  equation into Eq. (\ref{dress-charge-equation-2}) to readily obtain
\begin{align}
Z_{ub}(k^b) \approx & - \int_{-Q^u}^{Q^u} \textmd{d}k^u \, \cos k^u \, a_1(\sin k^u-k^b) - \int_{-Q^b}^{Q^b} \textmd{d}\tilde{k}^b \, a_2(k^b-\tilde{k}^b) \, Z_{ub}(\tilde{k}^b)\nonumber \\
\approx& -\frac{1}{\pi} \arctan \left.\left(\frac{\sin k^u-k^b}{|u|} \right)\right|_{k^u=-Q^u}^{k^u=Q^u} - \int_{-Q^b}^{Q^b} \textmd{d}\tilde{k}^b \, a_2(k^b-\tilde{k}^b) \, Z_{ub}(\tilde{k}^b) \notag\\
\approx & -\frac{2 Q^u}{\pi |u|}.
\end{align}
Similarly, we have
\begin{align}
Z_{bu}(k^u) \approx & -\int_{-Q^b}^{Q^b} \textmd{d}k^b \, a_1(k^b-\sin k^u) \nonumber \\
\approx & -\frac{1}{\pi} \arctan \left. \left( \frac{k^b-\sin k^u}{|u|} \right) \right|_{k^b=-Q^b}^{k^b=Q^b}\approx  -\frac{2 Q^b}{\pi |u|}, \\
Z_{bb}(k^b) \approx & 1-\int_{-Q^b}^{Q^b} \textmd{d}\tilde{k}^b \, a_2(k^b-\tilde{k}^b) \nonumber\\
\approx & 1-\frac{1}{\pi} \arctan \left.\left( \frac{k^b-\tilde{k}^b}{2|u|} \right)\right|_{\tilde{k}^b=-Q^b}^{\tilde{k}^b=Q^b} \approx  1-\frac{Q^b}{\pi |u|}.
\end{align}
Thus we obtain the dressed charge matrix to leading order, namely
\begin{align}\label{dress-charge-matrix1}
\hat{Z} \approx \left[
                  \begin{array}{cc}
                    1 & -\frac{2 Q^u}{\pi |u|} \\
                    -\frac{2 Q^b}{\pi |u|} & 1-\frac{Q^b}{\pi |u|} \\
                  \end{array}
                \right].
\end{align}

\begin{figure}
\centering
\includegraphics[width=0.95\linewidth]{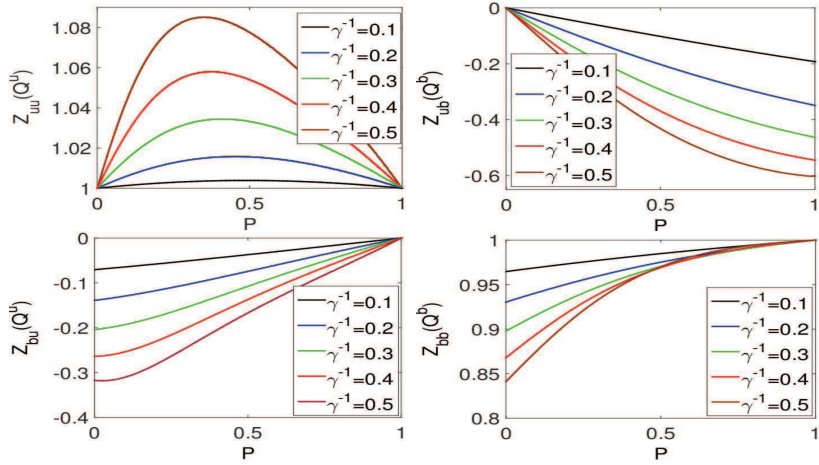}
\caption{Numerical results for the dressed charges $Z_{uu}(Q^u)$, $Z_{ub}(Q^b)$, $Z_{bu}(Q^u)$ and $Z_{bb}(Q^b)$  vs polarization  for different values of interaction strength.
Here we have denoted $\gamma=n/|u|$. The numerical solution is obtained by solving the dressed charge equations  (\ref{dress-charge-equation-1}-\ref{dress-charge-equation-4}). }
\label{dressed-charge-numeric}
\end{figure}

By virtue of the TBA equations (\ref{new0tba1}) and (\ref{new0tba2}) with the condition $\varepsilon^\gamma(\pm Q^\gamma)=0$ ($\gamma=u,b$),
and using  standard thermodynamic relations,  one can express the cut-off quasi-momenta  in terms of particle densities \cite{SC}
\begin{align}
Q^u \approx & \, \pi n^u+ \frac{2\pi}{|u|} \, n^u \, n^b, \label{fermi-point-u}\\
Q^b \approx & \, \frac{\pi}{\beta} n^b + \frac{\pi}{\beta^2 |u|} n^b (n^b+2n^u), \label{fermi-point-b}
\end{align}
where the density $n^\gamma={N^\gamma}/L$ ($\gamma=u,b$) must satisfy both conditions $n^\gamma/(\beta |u|) \ll 1$ and $n^\gamma/|u| \ll 1$.
In the above equations, the lattice  parameter is defined by
\begin{align}\label{lattice-effect-alpha}
\beta=\int_{-\pi}^\pi \textmd{d}k \, a_1\left(\sin k\right).
\end{align}
Furthermore, in the strong coupling regime, we have  $\beta \approx {2}/{|u|}$.
Without losing generality, the approximation used here requires that $n^\gamma$ ($\gamma=u,b$) is less than the order of $1/|u|$.
Meanwhile the condition  $n^\gamma/(\beta |u|)  \ll 1$ is required.
For the weak coupling regime, numerical calculation enables the confirmation of the asymptotic behaviour of the correlation functions.
In Fig.~\ref{dressed-charge-numeric}, we show the numerical solution of the dressed charge equations (\ref{dress-charge-equation-1})-(\ref{dress-charge-equation-4}).

We then substitute Eqs. (\ref{fermi-point-u}) and (\ref{fermi-point-b}) into Eq. (\ref{dress-charge-matrix1}).
Using the leading order  of $n^\gamma$ ($\gamma=u,b$) gives the dressed charge in the form
\begin{align}\label{dress-charge-matrix}
\hat{Z} \approx \left[
                  \begin{array}{cc}
                    1 & -\frac{2 n^u}{|u|} \\
                    -\frac{2 n^b}{|u|\beta} & 1-\frac{n^b}{\beta |u|} \\
                  \end{array}
                \right].
\end{align}
With the help of this dressed charge matrix,  the conformal dimensions given in Eqs.(\ref{cfd-1}) and (\ref{cfd-2}) in the low density regime can be approximated as
\begin{align}
2\Delta_\pm^u \approx & \left( \Delta D^u \pm \frac{1}{2}\Delta N^u  \right)^2 + 2\left( \Delta D^u \pm \frac{1}{2}\Delta N^u \right)\left( - \frac{2\, n^b}{|u|\beta} \Delta D^b \pm \frac{n^u}{|u|}\Delta N^b \right) + 2N_\pm^u\\
2\Delta_\pm^b \approx & \left( \Delta D^b \pm \frac{1}{2}\Delta N^b  \right)^2 + 2\left( \Delta D^b \pm \frac{1}{2}\Delta N^b \right)\left\{ -\frac{2 \, n^u}{|u|}\Delta D^u + \frac{n^b}{|u|\beta} \left[ -\Delta D^b \right. \right. \notag\\
&\left.\left. \pm \left( \Delta N^u + \frac{1}{2} \Delta N^b \right) \right]\right\} + 2N_\pm^b.
\end{align}
These results provide a direct calculation of the asymptotics of the correlation functions.

\section{Asymptotic behavior of correlation functions at zero temperature}
\label{sec-CR}

We study four types of correlation functions.
These are the single-particle Green's function $G_\uparrow (x,t)=\langle \hat c^\dag_{x,\uparrow}(t) \hat c_{0,\uparrow}(0)\rangle$,
the charge density correlation function $G_{nn}(x,t)=\langle \hat n^\dag_x (t)\hat n^\dag_{0}(0)\rangle$,
the spin correlation function $G_z(x,t)=\langle \hat s^z_x(t) \hat s^z_0(0)\rangle$,
and the pair correlation function $G_p(x,t) = \langle \hat c^\dag_{x,\uparrow} ( t )c^\dag_{x,\downarrow}(t) \hat c_{0,\uparrow}(0) \hat c_{0,\downarrow} (0)\rangle$.
Here $n(x,t)=\hat c_{x,\uparrow}^\dag (t) \hat c_{x,\uparrow} (t) + \hat c_{x,\downarrow}^\dag (t) \hat c_{x,\downarrow} (t)$ and
$S^z(x,t)=\frac{1}{2} \big[ \hat c_{x,\uparrow}^\dag (t) \hat c_{x,\uparrow} (t) - \hat c_{x,\downarrow}^\dag (t) \hat c_{x,\downarrow} (t) \big]$.
Each of these correlation functions is accessible through choosing suitable quantum numbers $\Delta N^\gamma$ ($\gamma=u,b$) with respect to the low-lying excitations.

The single-particle Green's function decays exponentially if the magnetic field $B<B_{c1}$, for which the external field does not provide enough energy to break up bound pairs.
However, if $B_{c1}<B<B_{c2}$ then the excess unpaired fermions appear in this gapless phase.
In this regime every correlation function satisfies power-law decay \cite{Belavin1984,Blote1986,Affleck1986,Cardy1986,Cardy1986,Izergin1989}.
The single-particle Green's function is determined by the quantum numbers $(\Delta N^u, \Delta N^b)=(1,0)$, which results in $(\Delta D^{u},\Delta D^{b}) \in (\mathbb{Z}+1/2,\mathbb{Z}+1/2)$.
Using Eq. (\ref{typical-correlation}) the leading terms of the single-particle Green's function are
\begin{align}
 G_\uparrow (x,t) \approx A_{\uparrow,1} \frac{\cos \left( \pi(n_\uparrow - 2 n_\downarrow) \right)}{|x+{\rm i} v^u t|^{2\theta_1} \,|x+{\rm i} v^b  t|^{2\theta_2}} +
 A_{\uparrow,2} \frac{\cos \left( \pi n_\uparrow \right)}{|x+{\rm i} v^u t|^{2\theta_3} \,|x+{\rm i} v^b t|^{2\theta_4}},
\end{align}
where the critical exponents are given by
\begin{align}
 \theta_1 \approx 1+\frac{2\, n^{b}}{|u|\beta}, \quad
 \theta_2 \approx \frac{1}{2}+\frac{2\, n^{u}}{|u|}-\frac{n^{b}}{|u|\beta}, \quad
 \theta_3 \approx 1-\frac{2\, n^{b}}{|u|\beta}, \quad
 \theta_4 \approx \frac{1}{2}-\frac{n^{u}}{|u|}-\frac{n^{b}}{|u|\beta}.
\end{align}
The leading order term is associated with the quantum numbers $(\Delta D^{u},\Delta D^{b})=\pm (1/2, -1/2)$,
with the next term coming from $(\Delta D^{u},\Delta D^{b})=\pm(1/2, 1/2)$.
The coefficients $A_{\uparrow,1}$ and $A_{\uparrow,2}$ cannot be derived from the  CFT approach,
yet this does not impede our understanding of the long-distance asymptotic behavior of the correlation functions.
Here we have introduced particle densities $n_\uparrow=(N^{u}+N^{b})/{L}$ and $n_\downarrow={N^{b}}/{L}$ for the particles with up and down spins.

We now turn to the charge density correlation function and the spin correlation function,
both of which are characterized by quantum numbers $(\Delta N^{u},\Delta N^{b})=(0,0)$,
implying $(\Delta D^{u},\Delta D^{b}) \in (\mathbb{Z},\mathbb{Z})$. The leading terms are expressed as
\begin{align}
 G_{nn} \approx & \, n^2 + A_{nn,1} \frac{\cos \left( 2\pi(n_\uparrow-n_\downarrow)x \right)}{|x+{\rm i} \, v^u t|^{2\theta_1}}
 +A_{nn,2} \frac{\cos \left( 2\pi n_\downarrow x \right)}{|x+{\rm i} \, v^b t|^{2\theta_2}} \notag\\
 &+A_{nn,3} \frac{\cos \left( 2\pi(n_\uparrow-2n_\downarrow)x \right)}{|x+{\rm i} \, v^u t|^{2\theta_3} \, |x+{\rm i} \, v^b t|^{2\theta_4}},\\
G_z
\approx & \, m_z^2 +  A_{z,1} \frac{\cos \left( 2\pi(n_\uparrow-n_\downarrow)x \right)}{|x+{\rm i} \, v^u t|^{2\theta_1}}
+ A_{z,2} \frac{\cos \left( 2\pi n_\downarrow x \right)}{|x+{\rm i} \, v^b  t|^{2\theta_2}} \notag\\
&+A_{z,3} \frac{\cos \left( 2\pi(n_\uparrow-2n_\downarrow)x \right)}{|x+{\rm i} \, v^u t|^{2\theta_3} \, |x+{\rm i} \, v^b t|^{2\theta_4}},\label{Sz-Sz}
\end{align}
where the critical exponents are given by
\begin{align}
\theta_1 \approx 2, \quad
\theta_2 \approx 2-\frac{4\, n^{b}}{|u|\beta}, \quad
\theta_3 \approx 2+\frac{8\, n^{b}}{|u|\beta}, \quad
\theta_4 \approx 2+\frac{8\,n^{u}}{|u|}-\frac{4\, n^{b}}{|u|\beta}.
\end{align}
Here the constant terms $n^2$ and $m_z^2$ originate from quantum numbers $(\Delta D^{u},\Delta D^{b})=(0,0)$, while the second, third, and fourth terms come
from $(\Delta D^{u},\Delta D^{b})=\pm (1,0)$, $\pm (0,1)$ and $\pm (-1,1)$, respectively.
The the amplitudes  $A_{nn,i}$ and $A_{z,i}$ ($i=1,2,3$) like $A_{\uparrow,1}$ and $A_{\uparrow,2}$ cannot be derived from the  CFT approach too.
For the first two oscillating terms in the above  correlators, the exponents are very close in low density limit. We thus have to consider the roles of amplitudes. 
By comparing  our analytical results with the numerical results given in \cite{Feiguin2007,Tezuka2008}, we  confirm the existence of the leading oscillation term with the frequency $2\pi (n_\uparrow-n_\downarrow)$.
They showed  that the dominant   contribution in the spin-spin correlation should  come from the term with spatial oscillation frequency  $2\pi(n_\uparrow-n_\downarrow)$, see Eq. (\ref{Sz-Sz}).

Last but not least we discuss the pair correlation function $G_p(x,t)$, which is described by the quantum numbers $(\Delta N^{u},\Delta N^{b})=(0,1)$,
allowing $(\Delta D^{u},\Delta D^{b}) \in (\mathbb{Z}+{1}/{2},\mathbb{Z})$.
We find that the leading terms for the pair correlation function are
\begin{align}
G_p(x,t)\approx &\, A_{p,1} \frac{\cos \left( \pi(n_\uparrow-n_\downarrow)x \right)}{|x+{\rm i} \, v^u  t|^{2\theta_1}\,|x+{\rm i} \, v^b  t|^{2\theta_2}}
+ A_{p,2} \frac{\cos \left( \pi(n_\uparrow-3n_\downarrow)x \right)}{|x+{\rm i} \, v^u  t|^{2\theta_3}\,|x+{\rm i} \, v^b  t|^{2\theta_4}},
\end{align}
with critical exponents
\begin{align}
\theta_1 \approx \frac{1}{2}, \quad
\theta_2 \approx \frac{1}{2}+\frac{n^{b}}{|u|\beta}, \quad
\theta_3 \approx \frac{1}{2}-\frac{4\, n^{b}}{|u|\beta}, \quad
\theta_4 \approx \frac{5}{2}-\frac{4\,n^{u}}{|u|}-\frac{3\, n^{b}}{|u|\beta}.
\end{align}
Here the first and the second terms are associated with $(\Delta D^u,\Delta D^b)=\pm (\frac{1}{2},0)$ and $\pm (\frac{1}{2},1)$, respectively.

The asymptotic behaviour of the correlation functions reveals an important many-body correlation nature and lattice effect,
which is apparent in the lattice parameter-dependent critical exponents.
Moreover, the leading order terms of the pair correlation function and spin correlation function reveal spatial oscillating behavior in their long-distance asymptotics.
The pair correlation function oscillates with a wave number $\Delta k = \pi (n_\uparrow-n_\downarrow)$.
So does the spin correlation function with $2\Delta k$.
Our analytic results provide a confirmation of the previous numerical observations of this oscillatory nature \cite{Feiguin2007,Luscher2008,Rizz2008,Tezuka2008,Tezuka2010}.
The oscillation stems from the backscattering processes in the two Fermi seas, where the  imbalance between the densities of spin-up and spin-down particles results in the mismatch of their Fermi surfaces.
It is interesting to see that the spatial oscillation in the 1D attractive Hubbard model is the feature of the Larkin-Ovchinnikov phase predicted in \cite{LO1965}.
We find that the oscillation terms in the spin and pair correlation functions arise from the Type III  elementary excitations (backscattering process).
Our theoretical result for the wave number shows good agreement with the numerics \cite{Feiguin2007,Tezuka2008},
where the numerical wave number is almost $\Delta k=\pi (n_\uparrow-n_\downarrow)$ for finite  $x, t$.
 Due to this reason,  $A_{z,1}$ might be much larger than $A_{z,2}$ and also $A_{z,3}$ for  for finite  $x, t$.
  The amplitude $A_{nn,i}$ and $A_{z,i}$ ($i=1,2,3$) cannot be derived through the conformal field theory approach. For the asymptotic behaviour of correlations, the amplitudes $A_{nn,i}$ and other amplitudes are not important in the limit $x, t\to \infty$.
Finally,  we point out that the parameter $\beta$ defined in Eq. (\ref{lattice-effect-alpha}) represents the lattice effect,
and thus distinguishes the Hubbard model from its continuum limit --  the 1D attractive SU(2) Fermi gas \cite{SC}.

\begin{figure}[t]
 \begin{center}
  \includegraphics[width=0.8\linewidth]{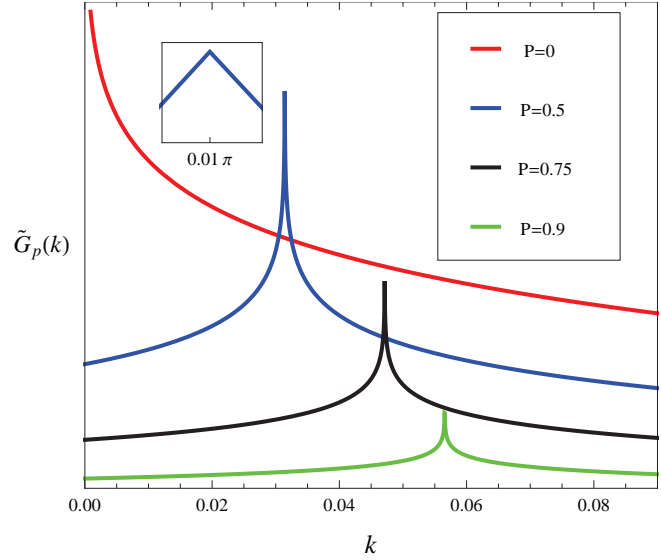}
 \end{center}
 \caption{The pair correlation function in momentum space $\tilde{G}_p(k)$ vs $k$ for different polarization $P=0, 0.5, 0.75$, and $0.9$ with interaction $u=-3$ and particle density $n=0.02$. The inset shows the singular behavior of $\tilde{G}_p(k)$ at $k=0.01\pi$ and $P=0.5$. This plot uses natural units for the quasimomentum $k$. See the  discussion on the amplitude of the pair correlation in text. }
 \label{f-pair}
\end{figure}

Applying Fourier transformation to the above correlation functions allows the derivation of their counterparts in momentum space \cite{Frahm1991PRB}.
For the equal-time correlation function,
\begin{align}
g(x,t=0^+)=\frac{\exp({\rm i} \, k_0  x)}{(x-{\rm i} \, 0)^{2\Delta_+} \, (x+{\rm i} \, 0)^{2\Delta_-}},
\end{align}
where $\Delta_\pm=\Delta_\pm^{u}+\Delta_\pm^{b}$. The Fourier transformation is thus
\begin{align}
\tilde{g}(k \approx k_0) \sim \left[\textmd{sign}(k-k_0) \right]^{2s} |k-k_0|^\nu.
\end{align}
Here the conformal spin and the exponent  are given by $s=\Delta_+ - \Delta_-$  and $\nu=2(\Delta_+ + \Delta_-)-1$.
Consequently, the Fourier transforms of the equal-time correlation functions near the singularities $k \approx k_0$ are expressed as
\begin{align}
\tilde{G}_\uparrow(k) \sim & \left[ \textmd{sign}\left(k-\pi(n_\uparrow-2n_\downarrow) \right) \label{FT-Green} \right]^{2s_\uparrow}|k-\pi(n_\uparrow-2n_\downarrow)|^{\nu_\uparrow}, \\
\tilde{G}_{nn}(k) \sim & \left[ \textmd{sign}\left(k-2\pi(n_\uparrow-n_\downarrow) \right) \right]^{2s_{nn}}|k-2\pi(n_\uparrow-n_\downarrow)|^{\nu_{nn}}, \label{FT-density} \\
\tilde{G}_z(k) \sim & \left[ \textmd{sign}\left(k-2\pi(n_\uparrow-n_\downarrow) \right) \right]^{2s_z}|k-2\pi(n_\uparrow-n_\downarrow)|^{\nu_z}, \label{FT-spin}\\
\tilde{G}_p(k) \sim & \left[ \textmd{sign}\left(k-\pi(n_\uparrow-n_\downarrow) \right) \right]^{2s_p}|k-\pi(n_\uparrow-n_\downarrow)|^{\nu_p},\label{FT-pair}
\end{align}
where
\begin{align}
2s_\uparrow \approx & \, 1, \quad
\nu_\uparrow \approx \, \frac{1}{2}+\frac{2\,n^{u}}{|u|}+\frac{n^{b}}{|u|\beta},\\
2s_z =&\, 2s_{nn} \approx  \, 0, \quad
\nu_z=\nu_{nn} \approx 1,\\
2s_p \approx & \, 0, \quad
\nu_p \approx \frac{n^{b}}{|u|\beta}.
\end{align}
We plot the pair correlation function in momentum space in Fig.~\ref{f-pair}.
This plot just demonstrates the peak positions and shows finite divergence of the peaks.
The amplitude of the pair correlation is not valid in this approach.
Therefore any $y$-axial scale is meaningless.

Notably, the correlation functions in momentum space shown in Eqs. (\ref{FT-Green}) to (\ref{FT-pair})
are valid only in the vicinity of the wave numbers $k_0$, i.e., $k \approx k_0$.
Fig. \ref{f-pair} shows that $\tilde{G}_p(k)$ has a singularity in non-zero momentum in a partially polarized phase.
This qualitatively agrees with the numerical result given  in \cite{Feiguin2007}.
 One should notice that this plot is correct  only if $k \approx \pi(n_\uparrow-n_\downarrow)$, where  the extrapolation is used  for the purpose of better visualization.

\section{Conclusion}
\label{sec-c}

In this paper  we have investigated four types of correlation functions for the 1D attractive Hubbard model at  zero temperature.
 The finite-size corrections to the ground state and the low-lying excitations are derived explicitly.
 Based on these  corrections to the momentum and energy, we have applied CFT in the study of long-distance asymptotic behavior of the correlation functions.
 The critical exponents have been obtained explicitly in this way.
 In contrast to the Fermi gas, these asymptotics of correlation functions essentially depend on the  parameter $\beta$ which represents the lattice effect.
 We have found that the spin and pair correlation functions have spatial oscillations with frequencies $2\pi(n_\uparrow-n_\downarrow)$ and $\pi(n_\uparrow-n_\downarrow)$, respectively.
 From the perspective of CFT, this type of oscillation is induced by the backscattering process of unpaired fermions and bound pairs among the Fermi points.
This gives a microscopic origin of the FFLO pair correlation  in the 1D system.
Using the Fourier transform, we have also derived  the correlation functions in momentum space.
The pair correlation is singular at the mismatch point $k=\pi(n_\uparrow-n_\downarrow)$, which confirms the frequency found by numerical methods  \cite{Feiguin2007}.
Meanwhile the correlation functions at  zero temperature display power-law decay in the partially polarized phase IV.
This suggests an analog of long-range order in a 1D many-body system and thus demonstrates the existence of a superconducting state.
We further point out that the dressed charge matrix in this CFT approach can be numerically resolved for arbitrary interaction strength $u<0$ (see \cref{dressed-charge-numeric}).
It follows that one can calculate the conformal dimensions and critical exponents for arbitrary interaction strength.
Our results provide benchmark physics of 1D strongly correlated fermions on a lattice which may be testable  in current ultracold atomic  experiments  \cite{Singha2011,Hart2015,Greif2016,Parsons2016S,Cheuk2016S,Boll2016S}.

\section*{Acknowledgement}
The authors  thank R. Hulet  for helpful discussion. This work is supported by
Key NNSFC grant number 11534014,   the National Key R\&D Program of China Grant Number 2017YFA0304500,
NNSFC grant numbers 11374331, 11174375 and ARC Discovery Projects DP130102839, DP170104934, DP160102337.

\appendix

\section{Finite-size corrections for the three types of excitations}
\label{sec-a-23}

In this Appendix we calculate the leading finite-size corrections for Type I, Type II and Type III excitations.

The situation for Type I elementary excitations is simple, which are a special case of particle-hole excitations.
 The change in energy of a particle-hole excitation is expressed by
\begin{align}
\Delta E=\sum_{\alpha=u,b} \sum_{\beta=+,-} \left[ \varepsilon^\alpha(k_{p,\beta}^\alpha)-\varepsilon^\alpha(k_{h,\beta}^\alpha) \right],
\end{align}
where we use subscripts $p$ and $h$ to label a particle and a hole.
Due to the Type I excitation taking place close to the Fermi points, the difference in the last equation can be approximated as the leading term of a Taylor expansion around the Fermi points.
Hence, the change in energy of Type I excitations is written as
\begin{align}
\Delta E \approx &\sum_{\alpha=u,b} \sum_{\beta=+,-} \left[ \left.\frac{\partial \varepsilon^\alpha}{\partial k^\alpha}\right|_{k^\alpha=Q_\beta^\alpha} \, \left( k_{p,\beta}^\alpha - k_{h,\beta}^\alpha \right)  \right]. \nonumber\\
\end{align}
Then noticing that the counting function connects the momentum $k^\alpha$ and quantum number $I^\alpha$, we similarly introduce a Taylor expansion in the last equation, with result
\begin{align}
\Delta E \approx &\sum_{\alpha=u,b} \sum_{\beta=+,-}\left[ \left.\frac{\partial \varepsilon^\alpha}{\partial k^\alpha}\right|_{k^\alpha=Q_\beta^\alpha} \, \left.\frac{\textmd{d}k^\alpha}{\textmd{d}y^\alpha}\right|_{k^\alpha=Q_\beta^\alpha} \frac{2\pi}{L} N_\beta^\alpha \, \textmd{sign}(\beta)   \right] \notag\\
=&\sum_{\alpha=u,b} \frac{2\pi}{L} v^\alpha \left( N_+^\alpha+N_-^\alpha \right),
\end{align}
where $y^\alpha= \lim_{L\rightarrow \infty} y_L^\alpha(k^\alpha)=p^\alpha(k^\alpha)$ is the counting function in the thermodynamic limit,
and we have used the definition
\begin{equation*}
v^\gamma=\pm \left.\frac{\textmd{d}\varepsilon^\gamma(k^\gamma)/\textmd{d}k^\gamma}{\textmd{d} p^\gamma (k^\gamma)/\textmd{d} k^\gamma}\right|_{k^\gamma=\pm Q^\gamma}
\end{equation*}
for the sound velocity.

In light of the zero momentum in the ground state, we can write down the change in total momentum,
\begin{align}
\Delta P=\sum_{\alpha=u,b} \frac{2\pi}{L} \left( N_+^\alpha-N_-^\alpha \right),
\end{align}
where $N_+^\alpha = I_{+,e}^\alpha - I _{-,g}^\alpha$ and
$N_-^\alpha = I_{-,g}^\alpha- I _{+,e}^\alpha$ arise from the change in distribution of quantum numbers close to the right and left Fermi points, respectively.
The subscripts $e$ and $g$ represent the excited state and ground state.

On the other hand,  the derivations for the changes in energy and total momentum of Type II and III elementary excitations are rather complicated.
We give an outline of these calculations here.

Prior to discussion of Type II and III excitations, we write down the root density in the thermodynamic limit,
\begin{align}\label{atr-random1}
\rho^\gamma(k^\gamma)=\rho_0^\gamma(k^\gamma)+ \sum_{\beta=u,b} \int_{Q^\beta_-}^{Q^\beta_+} \textmd{d}k^\beta \, \hat{K}_{\gamma \beta}(k^\gamma;k^\beta) \rho^\beta(k^\beta),
\end{align}
where for simplicity we have denoted $\rho^u_0(k^u)=\frac{1}{2\pi}$, $\rho^b_0(k^b)=\frac{1}{2\pi} \int_{-\pi}^\pi \textmd{d}k \, a_1(k^b-\sin k)$,
and introduce the matrix expression
\begin{align}
\hat{K}\left( k^u,k^b;k,\Lambda \right)=&
\left[
\begin{array}{cc}
0 & \hat{K}_{ub}(k^u;\Lambda)\\
\hat{K}_{bu}(k^b;k) & \hat{K}_{bb}(k^b;\Lambda)\\
\end{array}
\right]\\
=&
\left[
\begin{array}{cc}
0 & -\cos k^u \, a_1(\sin k^u-\Lambda) \\
-a_1(k^b-\sin k) & -a_2(k^b-\Lambda) \\
\end{array}
\right]
\end{align}
for the integral kernels. We then use a simpler expression for Eq. (\ref{atr-random1}),
\begin{align}\label{root-dens-uni}
\rho^\gamma(k^\gamma)=\rho_0^\gamma(k^\gamma)+\sum_{\beta=u,b}\hat{K}_{\gamma \beta}(k^\gamma;k^\beta) \otimes \rho^\beta(k^\beta),
\end{align}
where $\otimes$ stands for an integral on $k^\beta$ over interval $[Q^\beta_-,Q^\beta_+]$.

The low-lying excitation corresponds to small variations in the Fermi points from the ground state case.
This allows expansion of the energy per site with respect to the changes of the Fermi points to leading order,
\begin{align}\label{expansion1}
&e_E(Q_\pm^u,Q_\pm^b) \notag\\
=&e_E(\pm Q^u,\pm Q^b) + \frac{1}{2}\sum_{\alpha=u,b} \left[ \left.\frac{\partial^2 e_E}{\partial (Q_+^\alpha)^2}\right|_G \,(Q_+^\alpha-Q^\alpha)^2 + \left.\frac{\partial^2 e_E}{\partial (Q_-^\alpha)^2}\right|_G \,(Q_-^\alpha+Q^\alpha)^2 \right],
\end{align}
where $e_E=\sum_{\alpha=u,b} \int_{Q_-^\alpha}^{Q_+^\alpha} \textmd{d}k^\alpha \, \varepsilon^\alpha(k^\alpha) \rho_0^\alpha(k^\alpha)$ is the energy per site, and
$\pm Q^\alpha$ denotes the ground state Fermi points. The subscript $G$ implies $Q_\pm^u=\pm Q^u$ and $Q_\pm^b=\pm Q^b$.
In this expansion, the first order derivative $\left.\frac{\partial e_E}{\partial (Q_+^\alpha)}\right|_G$ vanishes due to the fact that the Fermi points minimize the energy in the ground state.

With the help of the TBA equations and integral equations of root densities, the second order derivative is given by
\begin{align}
\left. \frac{\partial^2 e_E}{\partial (Q_\pm^\alpha)^2} \right|_G=\pm \left.\frac{\partial \varepsilon^\alpha(k^\alpha)}{\partial Q_\pm^\alpha}\right|_{G,k^\alpha=\pm Q^\alpha} \rho^\alpha(\pm Q^\alpha).
\end{align}
Substituting this result and the sound velocity
$v^\gamma=\pm \left.\frac{\textmd{d}\varepsilon^\gamma(k^\gamma)/\textmd{d}k^\gamma}{2\pi \rho^\gamma(k^\gamma)}\right|_G$ into Eq. (\ref{expansion1}) yields
\begin{align}
e_E(Q_\pm^u,Q_\pm^b) = &\, e_E(\pm Q^u,\pm Q^b)\nonumber\\
& + \pi\sum_{\alpha=u,b} v^\alpha \left[\rho^\alpha(Q^\alpha)\right]^2 \left[ (Q_+^\alpha-Q^\alpha)^2 + (Q_-^\alpha+Q^\alpha)^2 \right], \label{new-temp-17}
\end{align}
with additional higher order correction terms.

In order to rewrite $Q_+^\gamma-Q^\gamma$ and $Q_-^\gamma+Q^\gamma$ in terms of $\Delta N^\gamma$ and $\Delta D^\gamma$,
we introduce the notation $\nu^\gamma=\frac{I_+^\gamma-I_-^\gamma}{L}=\frac{N^\gamma}{L}$ and $\delta^\gamma=\frac{I_+^\gamma+I_-^\gamma}{2L}=\frac{D^\gamma}{L}$,
where $N^u$ ($N^b$) stands for the number of unpaired fermions (bound pairs). $D^\gamma$ is the position of the center of the Fermi sea for unpaired fermions (bound pairs).
Their explicit expressions are given by
\begin{align}
\nu^\gamma=&\int_{Q_-^\gamma}^{Q_+^\gamma} \textmd{d}k \, \rho^\gamma(k) \qquad (\gamma=u,b) \\
\delta^u=&\frac{1}{2} \left( \int_\pi^{Q_+^u} + \int_{-\pi}^{Q_-^u} \right)\textmd{d}k \, \rho^u(k) + \frac{1}{2\pi}\int_{Q_-^b}^{Q_+^b} \textmd{d}k^b \, \theta\left(\frac{k^b}{|u|}\right) \rho^b(k^b),\\
\delta^b=&\frac{1}{2} \left( \int_\infty^{Q_+^b} + \int_{-\infty}^{Q_-^b} \right) \textmd{d}k \, \rho^b(k),
\end{align}
with $\theta(x)=2\arctan(x)$.

The total differential of $\nu^\gamma$ ($\gamma=u,b$) with respect to $Q_\pm^\beta$ ($\beta=u,b$) in the vicinity of the ground state is
\begin{align}\label{atr-random2}
\textmd{d}\nu^\gamma=\sum_{\beta=u,b} \left[ \left.\frac{\partial \nu^\gamma}{\partial Q_+^\beta} \right|_G \textmd{d}Q_+^\beta + \left.\frac{\partial \nu^\gamma}{\partial Q_-^\beta} \right|_G \textmd{d}Q_-^\beta  \right].
\end{align}
We furthermore denote
\begin{align}
\textmd{d}\vec{\nu}=\left[
                      \begin{array}{c}
                        \textmd{d}\nu^u \\
                        \textmd{d}\nu^b \\
                      \end{array}
                    \right],
\quad
\textmd{d}\vec{Q}_\pm=\left[
                      \begin{array}{c}
                        \textmd{d}Q_\pm^u \\
                        \textmd{d}Q_\pm^b \\
                      \end{array}
                    \right],
\quad
\left\{ \hat{V}_\pm \right\}_{\gamma \beta}=\left.\frac{\partial \nu^\gamma}{\partial Q_\pm^\beta}\right|_G.
\end{align}
Then Eq. (\ref{atr-random2}) can be rewritten as a vector equation,
\begin{align}\label{full-deri-nu}
\textmd{d}\vec{\nu}=\hat{V}_+\textmd{d}\vec{Q}_+ + \hat{V}_-\textmd{d}\vec{Q}_-.
\end{align}
Here we need to calculate
\begin{align}\label{new-temp-11}
\frac{\partial \nu^\gamma}{\partial Q_\pm^\beta}=\int_{Q_-^\gamma}^{Q_+^\gamma} \textmd{d}k \, \frac{\partial \rho^\gamma (k)}{\partial Q_\pm^\beta} \pm \rho^\gamma(Q_\pm^\beta) \delta_{\gamma\beta}.
\end{align}
According to Eq. (\ref{root-dens-uni}), it is straightforward to derive
\begin{align}\label{new-temp-14}
\frac{\partial \rho^\gamma(k^\gamma)}{\partial Q_\pm^\beta}=\pm \hat{K}_{\gamma\beta}(k^\gamma;k^\beta=Q_\pm^\beta) \rho^\beta(Q_\pm^\beta)+\sum_{\eta=u,b} \hat{K}_{\gamma\eta}(k^\gamma;k^\eta)\otimes \frac{\partial \rho^\eta(k^\eta)}{\partial Q_\pm^\beta},
\end{align}
which together with the dressed charge equations yields
\begin{align}\label{atr-random3}
\frac{\partial \nu^\gamma}{\partial Q_\pm^\beta}=\pm \rho^\beta(Q_\pm^\beta) \, Z_{\gamma\beta}(k^\beta=Q_\pm^\beta).
\end{align}

We also introduce the notation
\begin{align}
\hat{Z}=&\left.\left[
              \begin{array}{cc}
                Z_{uu}(k^u=Q_+^u) & Z_{ub}(k^b=Q_+^b) \\
                Z_{bu}(k^u=Q_+^u) & Z_{bb}(k^b=Q_+^b) \\
              \end{array}
            \right]\right|_G,
\\
\hat{\rho}=&\left.\left[
              \begin{array}{cc}
                \rho^u(k^u=Q_+^u) & 0 \\
                0 & \rho^b(k^b=Q_+^b) \\
              \end{array}
            \right]\right|_G,
\end{align}
making use of which in Eq. (\ref{atr-random3}) gives the compact  expression
\begin{align}
\hat{V}_\pm=\pm \hat{Z} \cdot \hat{\rho}.
\end{align}
Now we can express Eq. (\ref{full-deri-nu}) as
\begin{align}
\textmd{d}\vec{\nu}=\hat{Z} \cdot \vec{\rho} \cdot \textmd{d}\vec{Q}_+ - \hat{Z} \cdot \vec{\rho} \cdot \textmd{d}\vec{Q}_-.
\end{align}

In the next stage, we similarly consider the total differential of $\delta^\gamma$ with respect to $Q_\pm^\beta$ in the vicinity of the ground state,
\begin{align}\label{atr-random4}
\textmd{d}\delta^\gamma=\sum_{\beta=u,b} \left[ \left.\frac{\partial \delta^\gamma}{\partial Q_+^\beta} \right|_G \textmd{d}Q_+^\beta + \left.\frac{\partial \delta^\gamma}{\partial Q_-^\beta} \right|_G \textmd{d}Q_-^\beta  \right].
\end{align}
Similarly, we introduce the notation
\begin{align}
\textmd{d}\vec{\delta}=\left[
                      \begin{array}{c}
                        \textmd{d}\delta^u \\
                        \textmd{d}\delta^b \\
                      \end{array}
                    \right],
\quad
\left\{ \hat{W}_\pm \right\}_{\gamma \beta}=\left.\frac{\partial \delta^\gamma}{\partial Q_\pm^\beta}\right|_G,
\end{align}
in terms of which Eq. (\ref{atr-random4}) can be rewritten as the vector equation
\begin{align}\label{full-deri-delta}
\textmd{d}\vec{\delta}=\hat{W}_+\textmd{d}\vec{Q}_+ + \hat{W}_-\textmd{d}\vec{Q}_-.
\end{align}
We further calculate
\begin{align}
\frac{\partial \delta^u}{\partial Q_\pm^\beta}=& \, \frac{1}{2} \rho^u(Q_\pm^u) \cdot \delta_{u\beta} \pm \frac{1}{2} \, \theta\left( \frac{Q_\pm^b}{|u|} \right) \, \rho^b(Q_\pm^b) \cdot \delta_{b\beta}\nonumber\\
&
+ \frac{1}{2} \left( \int_\pi^{Q_+^u}+\int_{-\pi}^{Q_-^u} \right) \textmd{d}k^u \, \frac{\partial \rho^u(k^u)}{\partial Q_\pm^\beta}
+ \frac{1}{2\pi} \int_{Q_-^b}^{Q_+^b} \textmd{d}k^b \, \theta \left( \frac{k^b}{|u|} \right) \frac{\partial \rho^b(k^b)}{\partial Q_\pm^\beta}
\label{new-temp-12} \\
\frac{\partial \delta^b}{\partial Q_\pm^\beta}=& \, \frac{1}{2}\rho^b(Q_\pm^b) \cdot \delta_{b\beta} +  \frac{1}{2} \left( \int_\infty^{Q_+^b}+\int_{-\infty}^{Q_-^b} \right) \textmd{d}k^b \, \frac{\partial \rho^b(k^b)}{\partial Q_\pm^\beta}.
\label{new-temp-13}
\end{align}

Substituting Eq. (\ref{new-temp-14}) into Eqs. (\ref{new-temp-12}) and (\ref{new-temp-13}) then yields
\begin{align}
\frac{\partial \delta^u}{\partial Q_\pm^\beta}=& \,\pm \frac{1}{2} \rho^\beta(Q_\pm^\beta) \left[ \left( \int_\pi^{Q_+^u}+\int_{-\pi}^{Q_-^u} \right) \textmd{d}k^u \, \hat{K}_{u\beta}(k^u;k^\beta=Q_\pm^\beta) \pm \delta_{u\beta} +\frac{1}{\pi} \theta\left( \frac{Q_\pm^b}{|u|} \right) \cdot \delta_{b\beta} \right] \nonumber\\
&+\frac{1}{2\pi} \int_{Q_-^b}^{Q_+^b} \textmd{d}k^b \, \frac{\partial \rho^b(k^b)}{\partial Q_\pm^\beta} \left[ \theta \left( \frac{k^b}{|u|} \right) + \pi \left( \int_\pi^{Q_+^u}+\int_{-\pi}^{Q_-^u} \right) \textmd{d}k^u \, \hat{K}_{ub}(k^u;k^b) \right],\\
\frac{\partial \delta^b}{\partial Q_\pm^\beta}=& \,\pm \frac{1}{2}\rho^b(Q_\pm^b) \left[ \left( \int_\infty^{Q_+^b} + \int_{-\infty}^{Q_-^b} \right) \textmd{d}k^b \, \hat{K}_{b\beta}(k^b;k^\beta=Q_\pm^\beta) \pm \delta_{b\beta} \right] \nonumber\\
&+\frac{1}{2}\sum_{\gamma=u,b} \int_{Q_-^\gamma}^{Q_+^\gamma} \textmd{d}k^\gamma \, \frac{\partial \rho^\gamma(k^\gamma)}{\partial Q_\pm^\beta} \left[ \left( \int_\infty^{Q_+^b}+\int_{-\infty}^{Q_-^b} \right) \textmd{d}k^b \, \hat{K}_{b\gamma}(k^b;k^\gamma) \right].
\end{align}

Now we  introduce a set of new integral equations similar to the dressed charge equations, of the form
\begin{align}\label{dress-sigma1}
\sigma_{u\eta}(k^\eta)=& \, \delta_{b\eta} \cdot \left[ \frac{1}{\pi} \theta \left( \frac{k^b}{|u|} \right) + \left( \int_\pi^{Q_+^u}+\int_{-\pi}^{Q_-^u} \right) \textmd{d}k^u \, \hat{K}_{ub}(k^u;k^b) \right] \notag\\
&+\sum_{\gamma=u,b} \hat{K}_{\eta\gamma}^T(k^\eta;k^\gamma) \otimes \sigma_{u\gamma}(k^\gamma), \\
\sigma_{b\eta}(k^\eta)=& \left( \int_\infty^{Q_+^b} + \int_{-\infty}^{Q_-^b} \right) \textmd{d}\tilde{k}^b \, \hat{K}_{b\beta}\left(\tilde{k}^b;k^\eta\right)+\sum_{\gamma=u,b} \hat{K}_{\eta\gamma}^T \left(k^\eta;\tilde{k}^\gamma\right)\otimes \sigma_{b\gamma}\left(\tilde{k}^\gamma\right),\label{dress-sigma2}
\end{align}
where $\eta=u,b$ and the integral kernel matrix $\hat{K}^T$ is written as
\begin{align}\label{kernel-matrix-t}
\hat{K}^T(k^u,k^b;k,\Lambda)
=&
\left[
\begin{array}{cc}
0 & \hat{K}^T_{ub}(k^u;\Lambda)\\
\hat{K}^T(k^b;k) & \hat{K}^T(k^b;\Lambda)\\
\end{array}
\right] \notag\\
=&
\left[
                               \begin{array}{cc}
                                 0 & -a_1(\sin k^u-\Lambda) \\
                                 -\cos k \, a_1(k^b-\sin k) & -a_2(k^b-\Lambda) \\
                               \end{array}
                             \right].
\end{align}
With the help of the above integral equations for $\sigma_{\alpha\beta}(k^\beta)$ and  Eq. (\ref{new-temp-14}), we obtain
\begin{align}
\frac{\partial \delta^\alpha}{\partial Q_\pm^\beta}
=\pm\frac{1}{2} \rho^\beta(Q_\pm^\beta) \left[ \sigma_{\alpha\beta}(Q_\pm^\beta) \pm \delta_{\alpha\beta} \right],
\end{align}
and
\begin{align}\label{atr-random5}
\left\{ \hat{W} \right\}_\pm= \left. \left\{ \pm\frac{1}{2} \rho^\beta(Q_\pm^\beta) \left[ \sigma_{\alpha\beta}(Q_\pm^\beta) \pm \delta_{\alpha\beta} \right] \right\} \right|_G.
\end{align}
We now further define
\begin{align}
\hat{\sigma}=\left.\left[
                \begin{array}{cc}
                  \sigma_{uu}(Q_+^u) & \sigma_{ub}(Q_+^b) \\
                  \sigma_{bu}(Q_+^u) & \sigma_{bb}(Q_+^b) \\
                \end{array}
              \right]\right|_G,
\end{align}
in terms of which Eq. (\ref{atr-random5}) can be recast in the vector form
\begin{align}
\hat{W}_\pm=\frac{1}{2} \left( \hat{\sigma} + \hat{I} \right) \cdot \hat{\rho}.\label{new-temp-16}
\end{align}

Moreover, $\hat{\sigma}$ can be expressed in terms of $\hat{Z}$. Taking the derivative of Eqs. (\ref{dress-sigma1}) and (\ref{dress-sigma2}) with respect to their arguments, we then obtain
\begin{align}\label{simga'1}
\sigma'_{u\eta}(k^\eta)=& \, 2a_1(k^b)\, \delta_{b\eta}-\left. \hat{K}_{u\eta}^T(k^u;k^\eta) \right|_{k^u=\pi}^{k^u=Q_+^\pi} - \left. \hat{K}_{u\eta}^T(k^u;k^\eta) \right|_{k^u=-\pi}^{k^u=Q_-^\pi} \notag\\
& - \sum_{\gamma=u,b}\left. \left[ \hat{K}_{\gamma\eta}^T(k^\eta;k^\gamma) \, \sigma_{u\gamma}(k^\gamma) \right] \right|_{k^\gamma=Q_-^\gamma}^{k^\gamma=Q_+^\gamma}
+\sum_{\gamma=u,b} \hat{K}_{\eta\gamma}(k^\eta;k^\gamma) \otimes \sigma'_{u\gamma}(k^\gamma), \\
\sigma'_{b\eta}(k^\eta)=& -\left.\hat{K}_{b\eta}^T(k^\eta;k^b) \right|_{k^b=\infty}^{k^b=Q_+^b} -\left.\hat{K}_{b\eta}^T(k^\eta;k^b) \right|_{k^b=-\infty}^{k^b=Q_-^b} \notag\\
& - \sum_{\gamma=u,b}\left. \left[ \hat{K}_{\gamma\eta}^T(k^\eta;k^\gamma) \, \sigma_{b\gamma}(k^\gamma) \right] \right|_{k^\gamma=Q_-^\gamma}^{k^\gamma=Q_+^\gamma}
+\sum_{\gamma=u,b} \hat{K}_{\eta\gamma}(k^\eta;k^\gamma) \otimes \sigma'_{b\gamma}(k^\gamma).\label{sigma'2}
\end{align}

Taking the integral of $\sigma'_{u\eta}(k^\eta)$ with respect to $k^\eta$ over interval $[Q_-^\eta,Q_+^\eta]$ leads to
\begin{align}
\delta_{u\beta}-\hat{Z}_{u\beta}^T-\sum_{\gamma=u,b} \hat{\sigma}_{u\gamma} \cdot \hat{Z}_{\gamma\beta}^T=0.
\end{align}
Similarly, the integral of $\sigma'_{b\eta}(k^\eta)$ leads to
\begin{align}
\delta_{b\beta}-\hat{Z}_{b\beta}^T-\sum_{\gamma=u,b} \hat{\sigma}_{b\gamma} \cdot \hat{Z}_{\gamma\beta}^T=0.
\end{align}
We rewrite the above two equations in the vector form
\begin{align}
\hat{I}-\hat{Z}^T-\hat{\sigma}\cdot \hat{Z}^T=0.
\end{align}
Inserting this result into Eq. (\ref{new-temp-16}) then yields
\begin{align}
\hat{W}_\pm=\frac{1}{2}\left( \hat{Z}^T \right)^{-1} \cdot \hat{\rho}.
\end{align}
Using this last equation, the result given in Eq. (\ref{full-deri-delta}) can be expressed as
\begin{align}
\textmd{d}\vec{\delta}=\frac{1}{2}\left( \hat{Z}^T \right)^{-1} \cdot \hat{\rho} \cdot \textmd{d}\vec{Q}_+ - \frac{1}{2}\left( \hat{Z}^T \right)^{-1} \cdot \hat{\rho} \cdot \textmd{d}\vec{Q}_+.\label{full-deri-delta-new}
\end{align}

Now we rewrite Eqs. (\ref{full-deri-delta-new}) and (\ref{full-deri-nu}) in the form
\begin{align}
\left[
  \begin{array}{c}
    \textmd{d}\vec{\nu} \\
    \textmd{d}\vec{\delta} \\
  \end{array}
\right]
=
\left[
  \begin{array}{cc}
     \hat{Z} & -\hat{Z} \\
    \frac{1}{2}\left(\hat{Z}^T\right)^{-1} & \frac{1}{2}\left(\hat{Z}^T\right)^{-1} \\
  \end{array}
\right]
\cdot
\left[
  \begin{array}{c}
    \hat{\rho} \,\textmd{d}\vec{Q}_+ \\
    \hat{\rho} \, \textmd{d}\vec{Q}_- \\
  \end{array}
\right],
\end{align}
whose inverse is given by
\begin{align}
\left[
  \begin{array}{c}
    \hat{\rho} \,\textmd{d}\vec{Q}_+ \\
    \hat{\rho} \, \textmd{d}\vec{Q}_- \\
  \end{array}
\right]
=
\left[
  \begin{array}{cc}
     \frac{1}{2}\hat{Z}^{-1} &  \hat{Z}^T \\
    -\frac{1}{2}\hat{Z}^{-1} &  \hat{Z}^T \\
  \end{array}
\right]
\cdot
\left[
  \begin{array}{c}
    \textmd{d}\vec{\nu} \\
    \textmd{d}\vec{\delta} \\
  \end{array}
\right].
\end{align}

Finally, we express Eq. (\ref{new-temp-17}) in terms of $\textmd{d}Q^\gamma_\pm$ ($\gamma=u,b$), which gives, subject to higher order correction terms,
\begin{align}
&e_E(Q_\pm^u,Q_\pm^b)-e_E(\pm Q^u,\pm Q^b) \nonumber\\
=&\pi\sum_{\gamma=u,b} v^\gamma \left[\rho^\gamma(Q^\gamma)\right]^2 \left[ (Q_+^\gamma-Q^\gamma)^2 + (Q_-^\gamma+Q^\gamma)^2 \right]  \nonumber\\
=&\pi\sum_{\gamma=u,b} v^\gamma \left[\rho^\gamma(Q^\gamma)\right]^2 \left[ (\textmd{d}Q_+^\gamma)^2 + (\textmd{d}Q_-^\gamma)^2 \right].
\end{align}
In light of the definitions of $\hat{S}$, $\hat{\rho}$, $\Delta \vec{N}$, $\Delta \vec{D}$, with $\nu^\gamma={N^\gamma}/{L}$ and $\delta^\gamma={D^\gamma}/{L}$ ($\gamma=u,b$),
this last equation gives our final result,
\begin{align}
&e_E(Q_\pm^u,Q_\pm^b)-e_E(\pm Q^u,\pm Q^b) \nonumber\\
=&\frac{2\pi}{L^2}\left[ \frac{1}{4} \left( \Delta\vec{N} \right)^T \cdot \left( \hat{Z}^{-1} \right)^T \cdot \hat{S}_v \cdot \hat{Z}^{-1} \cdot \Delta\vec{N} + \left( \Delta\vec{D} \right)^T \cdot \hat{Z} \cdot \hat{S}_v \cdot \hat{Z}^{T} \cdot \Delta\vec{D} \right], \label{elem-exci-23-energy}
\end{align}
for the leading finite-size correction term.

The change in total momentum caused by Type II and III excitations is given by
\begin{align}
\Delta P=&\sum_{\gamma=u,b} \sum_{j=1}^{N^\gamma} \frac{2\pi I_j^\gamma}{L} \notag\\
=&\sum_{\gamma=u,b} \frac{2\pi}{L} \,\frac{1}{2} \left( I_+^\gamma-I_-^\gamma \right) \left( I_N^\gamma + I_1^\gamma \right) \nonumber\\
=&\, \frac{2\pi}{L} \sum_{\gamma=u,b} \Delta D^\gamma \cdot \left( N^\gamma+\Delta N^\gamma  \right). \label{elem-exci-23-momentum}
\end{align}

After the above lengthy calculations, the changes in the energy and total momentum of the three types of elementary excitations have been derived.
The change in energy shown in Eqs. (\ref{elem-exci-1-energy}) and (\ref{elem-exci-23-energy}) is summarized as Eq.(\ref{finite-s-c-energy-ex}).
The change in total momentum shown in Eqs. (\ref{elem-exci-1-momentum}) and (\ref{elem-exci-23-momentum}) is summarized as Eq. (\ref{finite-s-c-momentum-ex}).


\section*{References}

\begin{thebibliography}{10}
\expandafter\ifx\csname url\endcsname\relax
  \def\url#1{\texttt{#1}}\fi
\expandafter\ifx\csname urlprefix\endcsname\relax\def\urlprefix{URL }\fi
\expandafter\ifx\csname href\endcsname\relax
  \def\href#1#2{#2} \def\path#1{#1}\fi

\bibitem{Rase} M. Rasetti, The Hubbard Model Recent Results, World Scientific, 1991.

\bibitem{Monto}  A Montorsi, The Hubbard Model: A Collection Of Reprints, World Scientific, 1992.

\bibitem{Baer} D. Baeriswyl, D. K. Campbell, J. M. P. Carmelo, F. Guinea and E. Louis, The Hubbard Model-Its Physics and Mathematical Physics, Springer, 1995.

\bibitem{Ess} F. H. L. Essler, H. Frahm, F. G\"{o}hmann, A. Kl\"{u}mper and V. E. Korepin, The One-Dimensional Hubbard Model, Cambridge University Press, 2005.

\bibitem{Singha2011} A. Singha, M. Gibertini, B. Karmakar, S. Yuan, M. Polini, G. Vignale, M. I. Katsnelson, A. Pinczuk, L. N. Pfeiffer, K. W. West and V. Pellegrini, \href{http://science.sciencemag.org/content/332/6034/1176}{Two-Dimensional Mott-Hubbard Electrons in an Artificial Honeycomb Lattice}, Science  332 (2011) 1176.

\bibitem{Hart2015} R. A. Hart, P. M. Duarte, T.-L. Yang, X. Liu, T. Paiva, E. Khatami, R. T. Scalettar, N. Trivedi, D. A. Huse and R. G. Hulet, \href{http://www.nature.com/nature/journal/v519/n7542/full/nature14223.html?foxtrotcallback=true}{Observation of antiferromagnetic correlations in the Hubbard model with ultracold atoms}, Nature 519 (2015) 211.

\bibitem{Greif2016} D. Greif, M. F. Parsons, A. Mazurenko, C. S. Chiu, S. Blatt, F. Huber, G. Ji and M. Greiner,
\href{http://science.sciencemag.org/content/351/6276/953}{Site-resolved imaging of a fermionic Mott insulator}, Science 351 (2016) 953.

\bibitem{Parsons2016S}
M.~F. Parsons, A.~Mazurenko, C.~S. Chiu, G.~Ji, D.~Greif and  M.~Greiner,
  \href{http://science.sciencemag.org/content/353/6305/1253}{Site-resolved
  measurement of the spin-correlation function in the Fermi-Hubbard model},
  Science 353 (2016) 1253.

\bibitem{Cheuk2016S}
L.~W. Cheuk, M.~A. Nichols, K.~R. Lawrence, M.~Okan, H.~Zhang, E.~Khatami,
  N.~Trivedi, T.~Paiva, M.~Rigol and M.~W. Zwierlein,
  \href{http://science.sciencemag.org/content/353/6305/1260}{Observation of
  spatial charge and spin correlations in the 2D Fermi-Hubbard model}, Science
  353 (2016) 1260.

\bibitem{Boll2016S}
M.~Boll, T.~A. Hilker, G.~Salomon, A.~Omran, J.~Nespolo, L.~Pollet, I.~Bloch and
  C.~Gross, \href{http://science.sciencemag.org/content/353/6305/1257}{Spin-
  and density-resolved microscopy of antiferromagnetic correlations in
  Fermi-Hubbard chains}, Science 353 (2016) 1257.

\bibitem{Lieb1968PRL}
E.~H. Lieb and F.~Y. Wu, \href{}{Absence of mott transition in an exact solution of the
  short-range, one-band model in one dimension}, Phys. Rev. Lett. 20 (1968)
  1445.

\bibitem{Lieb2003PA}
E.~H. Lieb and F.~Wu,
  \href{http://www.sciencedirect.com/science/article/pii/S0378437102017855}{The one-dimensional Hubbard model: a reminiscence}, Phys. A 321 (2003) 1.

\bibitem{FF1964} P. Fulde and R. A. Ferrell, \href{https://journals.aps.org/pr/abstract/10.1103/PhysRev.135.A550}{Superconductivity in a strong spin-exchange field}, Phys. Rev. 135 (1964) A550.

\bibitem{LO1965} A. I. Larkin and Yu. N. Ovchinnikov, \href{}{Nonuniform state of superconductors}, Sov. Phys. JETP 20 (1965) 762.

\bibitem{Matsuda2007} Y. Matsuda and H. Shimahara, \href{}{Fulde-Ferrell-Larkin-Ovchinnikov state in heavy fermion superconductors}, J. Phys. Soc. Japan 76 (2007) 051005.

\bibitem{Ptok2017} A. Ptok, K. J. Kapcia, P. Piekarz and A. M. Ole\'{s}, \href{}{The \textit{ab initio} study of unconventional superconductivity in $\textrm{CeCoIn}_5$ and FeSe}, New J. Phys. 19 (2017) 063039.

\bibitem{Bogoliubov1988} N. M. Bogoliubov and V. E. Korepin, \href{}{Formation of Cooper pairs in the Hubbard model}, Mod. Phys. Lett. B 1 (1988) 349.

\bibitem{Bogoliubov1989} N. M. Bogoliubov and V. E. Korepin, \href{}{The role of quasi-one-dimensional structure in the high-$T_c$ superconductivity}, Int. J. Mod. Phys. B 3 (1989) 427.

\bibitem{Bogoliubov1990} N. M. Bogoliubov and V. E. Korepin, \href{}{Correlation functions of the one-dimensional Hubbard model}, Teor. i Mat. Fiz. 82 (3) (1990) 331.


\bibitem{Feiguin2007} A. E. Feiguin and F. Heidrich-Meisner, \href{}{Pairing states of a polarized Fermi gas trapped in a one-dimensional optical lattice}, Phys. Rev. B 76 (2007) 220508(R).


\bibitem{Luscher2008}  A. L\"{u}scher, R. M. Noack and A. M. L\"{a}uchli, \href{}{Fulde-Ferrell-Larkin-Ovchinnikov state in the one-dimensional attractive Hubbard model and its fingerprint in spatial noise correlations}, Phys. Rev. A 78 (2008) 013637.

\bibitem{Rizz2008} M. Rizzi, M. Polini, M. A. Cazalilla, M. P. Tosi and R. Fazio, \href{}{Fulde-Ferrell-Larkin-Ovchinnikov pairing in one-dimensional optical lattices}, Phys. Rev. B 77 (2008) 245105.

\bibitem{Tezuka2008} M. Tezuka and M. Ueda, \href{}{Density-matrix renormalization group study of trapped imbalanced Fermi condensates}, Phys. Rev. Lett. 100 (2008) 110403.

\bibitem{Tezuka2010} M. Tezuka and M. Ueda, \href{}{Ground states and dynamics of population-imbalanced Fermi condensates in one dimension}, New J. Phys. 12 (2010) 055029.

\bibitem{Batrouni2008} G. G. Batrouni, M. H. Huntley, V.G. Rousseau and R. T. Scalettar, \href{}{Exact numerical study of pair formation with imbalanced
    Fermion populations}, Phys. Rev. Lett. 100 (2008) 116405.


\bibitem{Wolak2010} M. J. Wolak, V. G. Rousseau, C. Miniatura, B. Gremaud, R. T. Scalettar and G. G. Batrouni,
\href{}{Finite-temperature quantum Monte Carlo study of the one-dimensional polarized Fermi gas}, Phys. Rev. A 82 (2010) 013614.


\bibitem{Lee2011} J. Y. Lee and X.-W. Guan, \href{}{Asymptotic correlation functions and FFLO signature for the one-dimensional attractive spin-1/2 Fermi gas}, Nucl. Phys. B 53 (2011) 125.



\bibitem{Guan:2000} X.-W. Guan,  {Algebraic Bethe ansatz for the one-dimensional Hubbard model with open boundaries}, J. Phys. A: Math. Gen. 33 (2000) 5391.

\bibitem{Zhou:1996} H.  Q. Zhou, {Quantum integrability for the one-dimensional Hubbard open chain}, Phys. Rev. B {54} (1996) 41.

\bibitem{Guan:1997} X.-W. Guan,  M. Wang and S.-D. Yang, {Lax pair and boundary K-matrices for the one-dimensional Hubbard model}, Nucl. Phys. B 485 (1997) 685.

\bibitem{Shiroishi:1997} M. Shiroishi and M. Wadati,  J. Phys. Soc. Japan 66 (1997) 2288.

\bibitem{Li:2014}Y.-Y. Li, J. Cao,  W.-L. Yang, K.-J. Shi and Y. Wang, {Exact solution of the one-dimensional Hubbard model with arbitrary boundary magnetic fields}, Nucl. Phys. B 879 (2014) 98.

\bibitem{Lee1988} K.-J.-B. Lee and P. Schlottmann, \href{}{Thermodynamic Bethe-ansatz equations for the Hubbard chain with an attractive interaction}, Phys. Rev. B 38 (1988) 11566.

\bibitem{Ess1994b} F. H. L. Essler and V. E. Korepin, \href{}{SU(2) $\times$ SU(2)- invariant scattering matrix of the Hubbard model}, Nucl. Phys. B 426 (1994) 505.

\bibitem{SC} S. Cheng, Y.-C. Yu, M. T. Batchelor and X.-W. Guan, \href{}{Universal thermodynamics of the one-dimensional attractive Hubbard model
}, arXiv:1708.07784.


\bibitem{Woyna1989} F. Woynarowich, \href{}{Finite-size effects in a non-half-filled Hubbard chain}, J. Phys. A 22 (1989) 4243.


\bibitem{Frahm1990PRB}
H.~Frahm and V.~E. Korepin,
  \href{https://link.aps.org/doi/10.1103/PhysRevB.42.10553}{Critical exponents for the one-dimensional hubbard model}, Phys. Rev. B 42 (1990) 10553.

\bibitem{Frahm1991PRB}
H.~Frahm and V.~E. Korepin,
  \href{http://link.aps.org/doi/10.1103/PhysRevB.43.5653}{Correlation functions of the one-dimensional hubbard model in a magnetic field}, Phys. Rev. B 43
  (1991) 5653.

\bibitem{Belavin1984}  A. A. Belavin, A. M. Polyakov and A. B. Zamolodchikov, \href{}{Infinite conformal symmetry in two-dimensional quantum field theory}, Nucl. Phys. B 241 (1984) 333.

\bibitem{Blote1986} H. W. Bl\"{o}te, J. L. Cardy and M. P. Nightingale, \href{}{Conformal invariance, the central charge, and universal finite-size
amplitudes at criticality}, Phys. Rev. Lett. 56 (1986) 742.

\bibitem{Affleck1986} I. Affleck, \href{}{Universal term in the free energy at a critical point and the conformal anomaly}, Phys. Rev. Lett. 56 (1986) 746.

\bibitem{Cardy1986} J. L. Cardy, \href{}{Operator content of two-dimensional conformally invariant theories}, Nucl. Phys. B 270 (1986) 168.

\bibitem{Izergin1989} A. G. Izergin, V. E. Korepin and N. Yu. Reshetikhin, \href{}{Conformal dimensions in Bethe ansatz solvable
models}, J. Phys. A: Math. Gen. 22 (1989) 2615.


\bibitem{Zvyagin:1993}A. A. Zvyagin, Sov. Phys. JETP 76 (1993) 167.

\end{thebibliography}
\end{document}